%% file: main.tex
\documentclass[letterpaper,twocolumn,10pt]{article}
\usepackage{arxiv}
\input{mydefs}

\begin{document}
\fancyhead{}

\title{\Large \bf \method: Differentially Private Graph Data Publication \\ by Exploiting Community Information}

\author{
{
\rm Quan Yuan\textsuperscript{1}}
\ \ \
{\rm Zhikun Zhang\textsuperscript{2,3}}
\ \ \ 
{\rm Linkang Du\textsuperscript{1}}
\ \ \
{\rm Min Chen\textsuperscript{2}}
\\
{\rm Peng Cheng\textsuperscript{1}}
\ \ \
{\rm Mingyang Sun\textsuperscript{1}\thanks{Mingyang Sun is the corresponding author.}}
\\ 
\\
{\rm \textsuperscript{1}\textit{Zhejiang University}} 
\\
{\rm \textsuperscript{2}\textit{CISPA Helmholtz Center for Information Security}} \\
{\rm \textsuperscript{3}\textit{Stanford University}} \\
}

\maketitle 

\begin{abstract}

Graph data is used in a wide range of applications, while analyzing graph data without protection is prone to privacy breach risks.
To mitigate the privacy risks, we resort to the standard technique of differential privacy to publish a synthetic graph.
However, existing differentially private graph synthesis approaches either introduce excessive noise by directly perturbing the adjacency matrix, or suffer significant information loss during the graph encoding process.
In this paper, we propose an effective graph synthesis algorithm \method by exploiting the community information. 
Concretely, \method differentially privately partitions the private graph into communities, extracts intra-community and inter-community information, and reconstructs the graph from the extracted graph information.
We validate the effectiveness of \method on six real-world graph datasets and seven commonly used graph metrics.

\end{abstract}

\section{Introduction}
\label{sec:introduction}

Many real-world systems can be represented by graphs, such as social networks~\cite{leskovec2010signed}, 
email networks~\cite{leskovec2007graph}, 
voting networks~\cite{mucha2010voting}, \etc,
and analyzing these graph data is beneficial in a wide range of applications~\cite{sharma2019brief}.
For instance, 
Facebook analyzes social networks and makes friend recommendations based on the connections (edges) between various users (nodes)~\cite{jiang2016big}.
Due to the sensitive nature of the graph data, it cannot be directly analyzed without protection.
A classical approach to analyzing the graph data while preserving privacy is anonymization, which removes the identification information of the nodes~\cite{wang2014high,cheng2010k}.
However, previous studies have shown that the anonymized graphs can be easily deanonymized by the attackers when they have some auxiliary information~\cite{ji2016graph,qian2017social}.

To overcome the drawback of the anonymization techniques, differential privacy (DP)~\cite{dwork2006calibrating,du2021ahead,zhang2018calm,WCZSCLLJ21}, a golden standard in the privacy community, has been applied to protect the privacy of graph data~\cite{sala2011pygmalion,xiao2014differentially}.
The core idea of DP is to guarantee that a single node/edge has a limited impact on the final output.
Most of the previous studies on differentially private graph analysis focus on designing tailored algorithms for specific graph analysis tasks,
such as degree distribution~\cite{hay2009accurate},
subgraph counts~\cite{sun2019analyzing},
and community discovery~\cite{ji2020community}.
Our paper, on the other hand, focuses on a more general paradigm, which publishes a synthetic graph that is semantically similar to the original graph while satisfying DP.
This paradigm is superior to the tailored algorithms in the sense that it enables arbitrary downstream graph data analysis tasks.

\mypara{Existing Solutions}
There are multiple existing studies focusing on publishing a synthetic graph under DP guarantee.
In~\cite{nguyen2015differentially}, Nguyen \etal proposed the \emph{Top-m Filter} (\tmf) method that directly perturbs the adjacency matrix of the original graph.
\tmf adds Laplace noise to each cell of the adjacency matrix and selects the top-$m$ cells from the noisy matrix as edges for the synthetic graph, 
where $m$ is the total number of edges in the original graph.
Note that directly perturbing the adjacency matrix introduces excessive noise; thus \tmf can only restore a few true edges when the privacy budget is small.
Chen \etal~\cite{chen2014correlated} designed the \emph{density-based exploration and reconstruction} (\der) method to perturb and reconstruct the adjacency matrix.
\der relabels the nodes to make edges concentrate on specific areas of the adjacency matrix, and then leverages a quadtree to calculate the density of the adjacency matrix.
To satisfy DP, the synthetic graph is reconstructed from the perturbed density. 
However, since the perturbation noise oftentimes overwhelms the true densities of the sparse areas, it is difficult for \der to maintain the structure of the original graph,
and the computational complexity of constructing a quadtree is large.
Different from \tmf and \der that directly perturb the adjacency matrix of the original graph, \hrg encodes the graph into a \emph{hierarchical random graph} (HRG)~\cite{clauset2008hierarchical} under DP, which reduces the strength of the noise.
However, \hrg needs a significant amount of time to build the HRG and suffers graph structure distortion.
Qin \etal~\cite{qin2017ldpgen} proposed to divide the nodes into multiple groups using $k$-means clustering.
However, the clustering accuracy under noise perturbation is usually low, which affects the accuracy of graph reconstruction.

\mypara{Our Proposal}
Existing methods either introduce excessive noise by directly perturbing the adjacency matrix,
or suffer substantial information loss in the process of encoding the graph data.
In this paper, we propose \method that exploits community information of the graph data to strike the trade-off between the perturbation noise and information loss. 

To avoid the large perturbation caused by directly adding noise to each cell of the adjacency matrix, \method leverages a community division mechanism to group all nodes into multiple communities and add noise to the communities instead of nodes. 
However, existing community detection algorithms do not satisfy DP. 
Therefore, we design a two-step division mechanism under DP guarantee, \ie, community initialization and community adjustment.
\method generates an initial community partition in community initialization and further tunes the nodes division in community adjustment.
The community aggregates more information than each node, resulting in higher robustness to noise perturbation. 

Based on the intuition that edges in a community are denser and edges between communities are sparser, we design two mechanisms to extract, perturb and reconstruct the edges of intra-community and inter-community separately, which preserve the structure information and suppress the noise simultaneously. 
Furthermore, we propose a post-processing procedure to maintain data fidelity.

\mypara{Evaluation}
We conduct experiments on six real-world graph datasets to illustrate the superiority of \method.
The experimental results show that \method outperforms the state-of-the-art methods for most of the metrics.
For instance, when the privacy budget is 1, 
for the modularity metric,
\method achieves 51.3\% lower relative error than that of \hrg on the Facebook dataset.
We also compare \method with tailored private methods optimized for specific graph analysis tasks.
Then, we conduct an ablation study on the hyper-parameters of \method and provide guidelines to select them.
We further illustrate the effectiveness of \method on a real-world application, \ie, influence maximization, which aims to find a small subset of nodes (seed nodes) in a graph that could maximize influence spread.
We observe that the seed nodes obtained using \method achieves up to 58.6\% higher influence spread than that of \der on the Facebook dataset.

\mypara{Contributions}
In summary, the main contributions of this paper are three-fold: 
\begin{itemize}
[itemsep=2pt,topsep=2pt,parsep=0pt]
    \item We take a deep look at existing solutions on differentially private graph synthesis, and identify their major drawbacks.
    
    \item We propose a practical method \method to generate a synthetic graph under DP.
    The general idea is to group the nodes in the graph by community information to avoid introducing excessive noise, and adopt different reconstruction approaches based on the characteristics of intra-community and inter-community to retain the graph structure. 
    
    \item We conduct extensive experiments and a real-world case study on multiple datasets and metrics to illustrate the effectiveness of \method.
    \method is open-sourced at \url{https://github.com/Privacy-Graph/PrivGraph}.
\end{itemize}

\section{Preliminaries}
\label{sec:preliminary}

\subsection{Differential Privacy}
\label{subsec:differential privacy}

Differential Privacy (DP)~\cite{dwork2006calibrating} was originated for the data privacy-protection scenarios, where a \textbf{trusted data curator} collects data from individual users, perturbs the aggregated results, and then publishes it.
Intuitively, DP guarantees that any single sample from the dataset has only a limited impact on the output. 
Formally, we can define DP as follows:

\begin{definition}[$\varepsilon$-Differential Privacy]
{An algorithm $\mathcal{A}$ satisfies $\varepsilon$-differential privacy ($\varepsilon$-DP), where $\varepsilon>0$, if and only if for any two neighboring datasets $D$ and $D^{\prime}$, we have} 
\begin{equation*}
    \forall T \subseteq Range (\mathcal{A}):\Pr {\mathcal{A}(D) \in T }
    \le e^{\varepsilon} \Pr {\mathcal{A}(D^{\prime}) \in T} 
\end{equation*}
where Range $(\mathcal{A})$ denotes the set of all possible outputs of the algorithm $\mathcal{A}$.
\end{definition}

We consider two datasets $D$ and $D^{\prime}$ to be \emph{neighbors}, denoted as $D\simeq D^{\prime}$, if and only if $D = D^{\prime} + r $ or $ D^{\prime} = D + r$, where $D + r$ stands for the dataset resulted from
adding the record $r$ to the dataset $D$.

\mypara{Laplace Mechanism}
Laplace mechanism (LM) 
satisfies the DP requirements by adding random Laplace noise to the aggregated results.
The magnitude of the noise depends on ${GS}_f$, \ie, \emph{global sensitivity}, 
\begin{equation*}
{GS}_f = \max_{D\simeq  D^{\prime} } {\parallel f(D) - f(D^{\prime}) \parallel }_1, 
\end{equation*}
where $f$ represents the aggregation function and $D$ (or $D'$) is the users' data. 
When $f$ outputs a scalar, the Laplace mechanism $\mathcal{A}$ is given below:
\begin{equation*}
    \mathcal{A}_f(D) = f(D) + \mathcal{L}\left(\frac{{GS}_f}{\varepsilon}\right), 
\end{equation*}
where $\mathcal{L}(\beta)$ stands for a random variable sampled from the Laplace distribution $\Pr {\mathcal{L}(\beta)=x}=\frac{1}{2\beta}e^{-\left | x \right | / \beta}$.
When $f$ outputs a vector, $\mathcal{A}$ adds independent samples of $\mathcal{L}(\beta)$ to each element of the vector.

\mypara{Exponential Mechanism}
Laplace mechanism (LM) applies to the scenario where the output of $f$ is a real value, while the output of Exponential Mechanism (EM)~\cite{mcsherry2007mechanism} is an item from a finite set.
EM samples more accurate answers with higher probabilities based on an exponential distribution. 
It takes the data $v$ as input and samples a possible output $o$ from the set $\mathcal{O}$ according to a quality function $q$.
The approach requires to design a quality function $q$
which takes as input the data $v$, a possible output $o$, and outputs a quality score. 
The global sensitivity of the quality function is defined as
\begin{equation*}
    {GS}_q = \max_{o} \max_{v \simeq  v^{\prime}} \left | q(v,o) - q(v^{\prime},o) \right|. 
\end{equation*}
$\mathcal{A}$ satisfies $\varepsilon$-differential privacy under the following equation. 
\begin{equation*}
    \Pr {\mathcal{A}_q(v)=o} = \frac{
    {\rm exp}\left( \frac{\varepsilon}{2{GS}_q}q\left(v,o \right) \right)}
    {{\textstyle \sum\limits_{o^{\prime} \in \mathcal{O}}}
    {\rm exp}
    \left(\frac{\varepsilon}{2{GS}_q}q(v,o^{\prime })
    \right)}
\end{equation*}

\mypara{Composition Properties of DP}
The following composition properties of DP are commonly used for building complex differentially private algorithms from simpler subroutines.

\begin{itemize}[itemsep=0pt,topsep=2pt,parsep=0pt]
    \item \mypara{Sequential Composition}
    Combining multiple subroutines that satisfy differential privacy for
    $\{\varepsilon_{1}, \cdots,\varepsilon_{k}\}$ results in a mechanism satisfying
    $\varepsilon$-differential privacy for $\varepsilon = { \sum_{i}\varepsilon_{i}} $.
    
    \item \mypara{Parallel Composition}
    Given $k$ algorithms working on disjoint subsets, 
    each satisfying DP for $\{\varepsilon_{1},\cdots,\varepsilon_{k}\}$, the result satisfies $\varepsilon$-differential privacy for $\varepsilon=\max\{\varepsilon_{i}\}$.
    
    \item \mypara{Post-processing}
    Given an $\varepsilon$-DP algorithm $\mathcal{A}$, releasing $g(\mathcal{A}(D))$ for any $g$ still satisfies $\varepsilon$-DP, \ie, post-processing an output of a differential private algorithm does not incur additional loss of privacy. 
\end{itemize}

\subsection{Differentially Private Graph Analysis}
\label{subsec:dp_graph}
The edges of a graph may contain very
sensitive information~\cite{ji2016graph},
such as social contacts, 
personal opinions, and private communication records~\cite{xiao2014differentially,chen2014correlated}. 
Edge-DP~\cite{hay2009accurate} provides rigorous theoretical guarantees to protect the privacy of these connections by limiting the impact of any edges in the graph on the output. 
As a result, it offers meaningful privacy protection in many applications~\cite{ chen2014correlated,nguyen2015differentially,xiao2014differentially}.

More specifically, given a graph $G=(V,E)$, an edge neighboring graph $G^{\prime}=(V^{\prime},E^{\prime})$ can be obtained by adding (or removing) an edge, where $V$ ($V^{\prime}$) is the set of nodes and $E$ ($E^{\prime}$) is the set of edges. 
From~\cite{hay2009accurate},
the system difference $a \oplus b$ is the sets of elements in either set $a$ or set $b$, but not in both, \ie, $a \oplus b = (a \cup b) \setminus (a \cap b)$. 
Hence, the definitions of edge neighboring graph and $\varepsilon$-edge DP are as follows.

\begin{definition}[Edge neighboring graph]
Given a graph $G=(V,E)$, a graph $G^{\prime}=(V^{\prime},E^{\prime})$ is an edge neighboring graph of $G$ if and only if $\left | V \oplus V^{\prime} \right | + \left | E \oplus E^{\prime} \right | = 1 $.
\end{definition}

\begin{definition}[$\varepsilon$-edge differential privacy]

An algorithm $\mathcal{A}$ satisfies $\varepsilon$-edge differential privacy ($\varepsilon$-edge DP), where $\varepsilon>0$. If and only if for any two edge neighboring graphs G and $G^{\prime}$,
\begin{equation*}
 \forall T \subseteq Range (\mathcal{A}): \Pr {\mathcal{A}(G) \in T}\le e^{\varepsilon}  \Pr {\mathcal{A}(G^{\prime}) \in T}, 
\end{equation*}
where 
\emph{$Range(\mathcal{A}) $} denotes the set of all possible outputs of 
$\mathcal{A}$.
\end{definition}

\mypara{Discussion}
Edge-DP guarantees that any edges in the graph have limited impacts on the final output, instead of deleting specific edges from the graph. 
As such, the attacker cannot infer the existence of any edges %
by observing the final output. 
The attacker can reconstruct the graph relying on auxiliary information such as some users are from the same class, 
but this is orthogonal to the privacy guarantee provided by DP. 
If the attacker has such auxiliary information, they can reconstruct these edges regardless of whether they are published. 

Edge-DP is well-suited for scenarios where the edges are independent of each other, such as email communication records. 
However, in cases where the edges in a graph are correlated and can be deduced from one another, such as in a friendship graph, 
edge-DP's guarantees may be insufficient while $k$-edge DP~\cite{hay2009accurate} can still provide meaningful privacy protection. 
In $k$-edge DP, graph $G$ and $G^{\prime}$ are neighbors if $\left | V \oplus V^{\prime} \right | + \left | E \oplus E^{\prime} \right | \le k$.

The mechanism satisfying the formal definition of edge-DP also satisfies the requirements for $k$-edge DP. 
However, it's important to note that achieving the same level of quality in published results using $k$-edge DP will require $k$ times more privacy budget than using edge-DP. 
For example, {$\mathcal{A}$} satisfies $\varepsilon$-edge DP for $k=1$ and $\varepsilon=1$, it also satisfies $k$-edge $\varepsilon$-DP for $k=10$ and $\varepsilon=10$.

\subsection{Community Detection}
\label{subsec:preliminary_community_detection}
Community detection is an effective method for discovering densely connected subnetworks in graph data.
It has a wide range of practical applications.
For instance, in social networks, 
community detection helps identify a group of users with similar interests.
A series of classical community detection algorithms have been proposed~\cite{blondel2008fast,rosvall2008maps,chen2022graph}.

\mypara{Louvain}
The Louvain~\cite{blondel2008fast} method is adopted frequently due to its computing efficiency and outstanding grouping effect.
The optimization goal of the Louvain method is to maximize the modularity~\cite{guimera2004modularity}, which measures the quality of the community division. 
The definition of modularity is as follows. 

\begin{equation}
\small
        {Q} =
        \sum_{C}\left[\frac{\sum_{in}}{2m}-\left(\frac{{\sum_{tot}}}{2m}\right)^2\right], 
        \label{eq:modularity}
\end{equation}
where $\sum_{in}$ is the sum of the weights of the edges inside the community $C$, 
$\sum_{tot}$ stands for the sum of the weights of the edges incident nodes in $C$, 
and $m$ represents the sum of the weights of all edges.

At the beginning, Louvain randomly initializes each node as a single-node community. 
Then, Louvain iterates the following two processes until the modularity converges.
Firstly, for each node in the graph, Louvain finds the largest gain in modularity by assigning the node to its neighbors' community. 
If there exists no positive gain, the node will stay in the original community.
This phase stops when a local maximum of the modularity is obtained, \ie no individual move can enlarge the modularity.
Then, Louvain merges the nodes in a community into a super-node and updates the weights of super-nodes. 
The weights between super-nodes are determined by the sum of the edges' weights between two communities. 
The inner weight of a super-node is the sum of the edges' weights inside the community. 
The purpose of the first step is to achieve modularity optimization and the second step is to complete the aggregation of nodes in the same community. 
According to~\cite{fortunato2007resolution}, the modularity-based approaches might lose the small communities during the modularity optimization process. 
However, the information extraction and graph reconstruction processes of \method can recover these small communities. 
In the information extraction process, \method leverages a degree sequence to encode the edges in a community, where the dense connections of small communities are still preserved. 
Then, \method decodes the degree sequence and rebuilds the small communities. 
The above analysis are further supported by the experimental results in \autoref{subsec:preserve_small_community}. 

In addition, time-scale is a hyper-parameter to adjust the resolution of community detection~\cite{lambiotte2008laplacian}. 
By integrating time-scale $t$, 
modularity optimization can reveal the community structures at different resolutions,
and the objective function is updated as follows. 
\small
\begin{equation}
    {Q^{*}} =
    \sum_{C}\left[\frac{\sum_{in}}{2m}t-\left(\frac{{\sum_{tot}}}{2m}\right)^2\right]
    , \nonumber 
    \label{eq:resolution}
\end{equation}
\normalsize
where $t$ is the resolution parameter,
and the meanings of other parameters are the same as those in~\autoref{eq:modularity}.
When $t=0$, each node occupies one community.
When $t=1$, the optimization goal is consistent with traditional modularity optimization.
When $t$ increases, the number of communities usually decreases.
We further analyze the impact of different time-scale settings in~\autoref{subsec:impact_resolution}. 

\mypara{Discussion}
Intuitively, clustering can partition nodes into groups like Louvain. 
However, compared to Louvain, it introduces new hyperparameters, such as the node's feature, the distance metric, and the group size, which make clustering more challenging to tune under the noise perturbation. 

\section{Problem Definition and Existing Solutions}
\label{sec:problem_definition}

\subsection{Problem Definition}
\label{subsec:problem_definition}

\mypara{Threat Model}
With full access to the published graph, the adversary’s goal is to infer whether an edge exists in the original graph. 
For example, given a synthetic email communication network,
the adversary aims to determine if there is an email connection between any two users. 

In this paper, we consider an undirected and unweighted graph $G=(V,E)$, where $V$ is the set of nodes and $E$ is the set of edges.
We are interested in the following problem: \textit{Given a graph $G$, how to generate a synthetic graph $G_s$ that shares similar graph properties with the original graph $G$ while satisfying edge-DP}.
The synthetic graph $G_s$ can be used for any downstream graph analysis tasks without privacy loss due to the post-processing property of DP. 
We summarize the frequently used mathematical notations in \autoref{table:math_notations}.

Following previous studies~\cite{chen2014correlated,xiao2014differentially,nguyen2015differentially}, we measure the similarity between $G_s$ and $G$ from five different aspects:
Community discovery, node information, degree distribution, path condition, 
and topology structure.
Concretely, the community discovery aims to detect the communities and reveal the structure of the graph,
the node information reflects the neighboring information of each node, 
the degree distribution reveals the overall connection density of the graph, 
the path condition reflects the connectivity of the graph, 
and the topology structure illustrates the level of node aggregation. 

\begin{table}[!t]
    \centering
    \caption{Summary of mathematical notations.}
    \label{table:math_notations}
    \vspace{-0.1cm}
    \footnotesize
    \setlength{\tabcolsep}{1.2em}
	\begin{tabular}{cc}
		\toprule
		\textbf{Notation} & \textbf{Description}  \\
		\midrule
		$G$ & Graph \\
		$\varepsilon$ & Privacy budget  \\
		$N$ & The number of nodes per initial  community \\
		${m_1}$ & The number of initial communities  \\
		
		${m_2}$ & The number of final communities  \\
		$w$ & The weight of the weighted graph \\
		
		$\mathbb{C}$ &  The set of community partitions \\
		$C$ & The community composed of nodes \\
		
		$d$& Degree sequence \\
		$v$& Edge vector  \\
	
		$s$& Subgraph \\
		\bottomrule
	\end{tabular}
\end{table}

\subsection{Existing Solutions}
\label{section:existing_solutions}

\mypara{TmF~\cite{nguyen2015differentially}}
It first perturbs the adjacency matrix of the original graph by adding Laplace noise to each cell. 
Then, \tmf chooses the top-$m$ noisy cells as edges in the perturbed adjacency matrix, 
where $m$ is obtained by adding Laplace noise to the true number of edges to satisfy DP. 
However, perturbing the whole cells in the adjacency matrix introduces excessive noise. 
Most true edges cannot be retained from the top-$m$ noisy cells, especially when $\varepsilon$ is small. 

Although the computational cost of \tmf is small, \ie, running in linear computational complexity, it faces several limitations: 
1) \tmf cannot capture the structure of the graph accurately since the selection of edges is random. 
2) With a fixed privacy budget, the utility of the synthetic graph will decrease as the number of nodes increases. 

\mypara{DER~\cite{chen2014correlated}} 
Similar to TmF~\cite{nguyen2015differentially}, \der also processes the adjacency matrix of the original graph. 
\der mainly consists of three steps: Node relabeling, dense region exploration, and edge reconstruction.
\der first makes the edges in the adjacency matrix clustered together in the node relabeling step. 
Then, \der divides the adjacency matrix into multiple blocks and estimates the density by a noisy quadtree.
Finally, \der reconstructs the edges in each block by exponential mechanism. 

\der has two main drawbacks: 
1) The time- and space-complexity of \der are quadratic, hindering its applications for large scale graphs. 
2) The original graph structure is hard to maintain in dense region exploration. 

\mypara{PrivHRG~\cite{xiao2014differentially}}
\hrg captures the graph structure by using a statistical HRG~\cite{clauset2008hierarchical}, which aims to suppress the noise strength. 
The likelihood of an HRG for a graph $G$ shows how plausible the HRG is to represent $G$.
An HRG with a higher likelihood can represent the structure of the original graph better than those with lower likelihoods.

The method first maps all nodes of a graph into a hierarchical structure and records connection probabilities between any pair of nodes in the graph.
Then, \hrg uses the Markov chain Monte Carlo (MCMC) method to obtain an HRG with high likelihood while satisfying edge-DP.
Finally, the edges are reconstructed based on the perturbed probabilities. 

However, \hrg faces two limitations in practice:
1) It is time- and space-consuming to sample a high-quality HRG, especially when the amount of nodes is large. 
2) The partial information of the graph is lost when constructing an HRG, which decays the accuracy of \hrg. 

\mypara{LDPGen~\cite{qin2017ldpgen}}
\gen is initially designed to generate a synthetic graph under local differential privacy (LDP). 
\gen first divides all nodes into two groups randomly.
Then, \gen modifies the partition of the nodes by $k$-means clustering according to the number of connected edges from the nodes to each group, where the number of clusters is a pre-defined value. 
Finally, the connected edges are reconstructed based on the grouping and connected information.
The idea of \gen can be ported to the DP setting. 
The main modification is in the $k$-means clustering, 
where the number of connected edges from the nodes to each group can be obtained directly by perturbing the true values in the central-DP setting, 
instead of using noisy vectors from the previous phase to estimate them in the local-DP setting.

Nevertheless, \gen faces the following limitations:
1) The space-complexity of \gen is high, which limits its applications to large graphs. 
2) \gen utilizes the preset number of clusters to encode and perturb all edges at the same granularity, which may introduce excessive noise to the groups with sparse connection. 

\section{Our Proposal: \method}
\label{sec:methodology}

\subsection{Motivation}
When publishing a graph under edge-DP, perturbing each edge in the adjacency matrix, \ie, in a fine-grained manner, could introduce excessive noise. 
If we encode the entire graph into nodes' degree distribution, \ie, in a coarse-grained manner, the compression leads to information loss in the graph structure. 
From~\cite{rosvall2008maps,porter2009communities}, 
a natural graph usually consists of communities, such as social networks~\cite{lusseau2004identifying}, 
biological networks~\cite{girvan2002community}, and voting networks~\cite{mucha2010voting}.
Considering the edges in the same community tend to have similar structures, we can use coarse-grained aggregation to alleviate the perturbing noise. 
For the edges among communities, they occupy a small part of total edges but in various structures, thus we can adopt fine-grained encoding to reduce the information loss. 
Therefore, the communities can be a basis for the desired granularity. 
According to this observation, we design a graph data publishing approach, called \method, which achieves outstanding data utility and rigorously satisfies DP. 

\subsection{Overview}
\label{section:overview}

\begin{figure*}[htbp]
\centering
\includegraphics[width=0.9\textwidth] {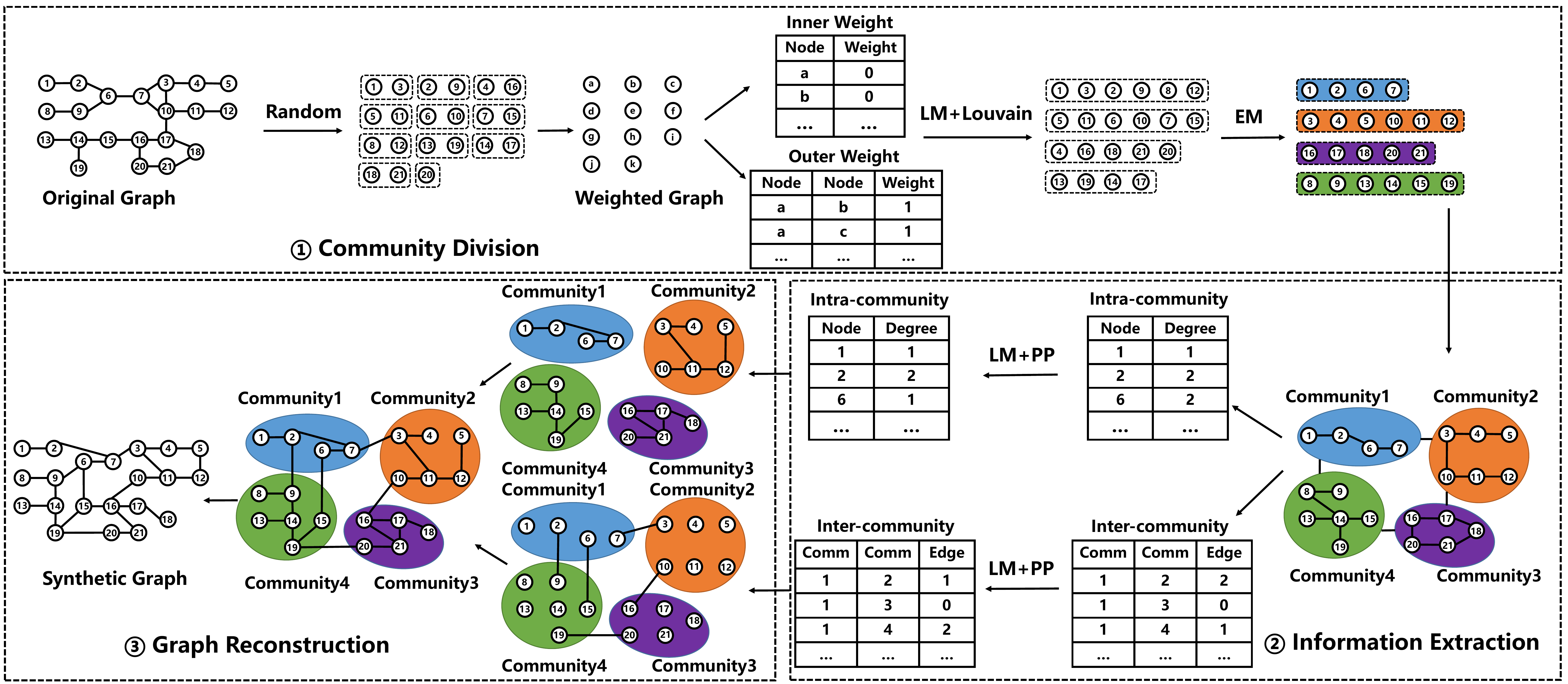}
\vspace{-0.3cm}
\caption{\method overview. \method is composed of three phases: Community division, information extraction, and graph reconstruction.
The community division phase includes two steps, \ie, community initialization and community adjustment.
The community initialization aims to obtain a preliminary community division by Laplace Mechanism and the Louvain method based on the random community division.
Then \method applies Exponential Mechanism to obtain the final community partitions in the community adjustment.
In the information extraction phase, 
\method calculates the degree sequences within communities and the number of edges between communities, and respectively perturbs them to satisfy DP. 
In the graph reconstruction phase, 
\method generates a synthetic graph based on the noisy information extracted from Phase 2. 
EM, LM, and PP stand for exponential mechanism, Laplace mechanism, and post-processing, respectively.}
\vspace{-0.3cm}
\label{fig:framework}
\end{figure*}

As shown in \autoref{fig:framework}, 
the workflow of \method consists three phases: Community division, information extraction, and graph reconstruction. 

\mypara{Phase 1: Community Division (CD)} 
We design a community detection algorithm to obtain a suitable nodes partition. 
The core idea is first to generate a coarse partition by merging several nodes into a super-node.
Then, as shown in the upper dashed box of \autoref{fig:framework}, the super-nodes form a weighted graph containing inner weight, \ie, the edges within a community, and outer weight, \ie, the edges among communities. 
\method separately perturbs the two parts by the Laplace noise, and applies post-processing to calibrate the noisy weights. 
Next, \method adopts the Louvain~\cite{blondel2008fast} method to refine the community division based on the calibrated weights. 
Finally, \method utilizes exponential mechanism to adjust and obtain the final partitions.
The details of Phase 1 are in~\autoref{section:community_division}. 

\mypara{Phase 2: Information Extraction (IE)} 
As shown in the bottom-right dashed box, we extract the information from the original graph based on the communities from Phase 1. 
Since most nodes tend to have more edges within communities and fewer edges between communities, 
we record each node's degree in their own communities and the sum of edges between community pairs. 
To satisfy the edge-DP, \method adds Laplace noise to the nodes' degrees and the sum of edges, and then conducts post-processing to the perturbed results. 
The details of Phase 2 are referred to~\autoref{section:information_extraction}. 

\mypara{Phase 3: Graph Reconstruction (GR)}
In the left-bottom dashed box, \method rebuilds the intra-community edges based on the noisy degree of each node. 
For the inter-community edges, \method randomly connects the nodes between different communities under the sum of edges constraints. 
The details of Phase 3 are in~\autoref{section:graph_construction}.

\subsection{Community Division} 
\label{section:community_division}
We divide the nodes into a number of communities by the following community division algorithm. 
For ease of exposition, we refer to the last  operation of Phase 1 in \autoref{section:overview} as \emph{community adjustment}. 
Therefore, there are two parts in Phase 1, \ie, obtaining the preliminary partitions (Community Initialization) and further adjustment based on exponential mechanism (Community Adjustment). 

\mypara{Community Initialization} 
As shown in~\autoref{algorithm:community_initial}, all nodes are divided into $m_1$ communities randomly at first. 
The purpose of node division is to reduce the dimension of original data and mitigate the noise perturbation.
Each community contains $N$ nodes except for the last community, 
which may not have exactly $N$ nodes. 
The nodes in the same community are considered as a super-node. 
Then, the original graph is converted to a weighted graph by the super-nodes. 
The sum of the nodes’ degrees within the same community is the inner weight of the super-node, 
and the number of edges between two communities is the outer weight of two super-nodes. 

In \autoref{algorithm:community_initial} (\autoref{line:weighted_graph_begin} - \autoref{line:weighted_graph_end}), 
\method adds Laplace noise to the weighted graph to guarantee edge-DP.
Note that the inner 
and outer
weights have different global sensitivities.
For inner weights, an edge effects the degrees of two nodes, \ie, $\Delta f_{i}=2$.
For outer weights, the number of edges between two communities changes 1 at most, \ie, $\Delta f_{o}=1$.

After adding Laplace noise, there may appear some negative weights. 
Thus, we adopt NormSub~\cite{wang2019consistent} to post-process the perturbed weights. 
Given the inner weighted vector $\tilde w$, we want to find an optimal integer $\delta^{*}=\mathop{\arg\min}\limits_{\delta}\left | \sum_{i \in D}\max(\tilde w_i + \delta,0)-\sum_{i \in D}\tilde w_i\right |$, where $D$ is the set of indexes of all inner weights.
Then, we iterate all elements of $\tilde w$ and update the value of $\hat w_i$ to $\max(\tilde w_i + \delta^{*},0)$. 
After the consistency processing, we obtain the inner weights
satisfying non-negative constraints. 
We also utilize NormSub to post-process the outer weights. 

Then, we leverage the Louvain method to further aggregate super-nodes into communities by maxing modularity. 
Since the weighted graph is already protected by DP, the Louvain method can be applied directly. 
The definition of modularity is shown in~\autoref{eq:modularity}. 
The Louvain method moves the node to the neighboring community with the highest modularity gain $\Delta Q$. 
The modularity gain $\Delta Q$ 
can be calculated by:
\begin{small}
\begin{equation*}
    \left[\frac{\sum_{in} + k_{n,in}}{2m}-\left(\frac{\sum_{tot}+k_n}{2m} \right)^2
    \right] - 
    \left[\frac{\sum_{in}}{2m}-\left(\frac{\sum_{tot}}{2m}\right)^2-\left(\frac{k_n}{2m}\right)^2
    \right],
\end{equation*}
\end{small}
where
$k_n$ stands for the sum of the weights of the edges incident to node $n$ and
$k_{n,in}$ stands for the sum of the weights of the edges from node $n$ to nodes in $C$ .

The nodes in the same community will be regarded as a super-node when the modularity no longer increases significantly. 
Through multiple rounds of iterations, the tightly connected super-nodes are merged to form new community partitions $\mathbb{C_W}$.

According to the correspondence between each super-node of the weighted graph and the nodes of the original graph, \ie, $\mathbb{C_R}$,
and the partitions of all super-nodes in the weighted graph, \ie, $\mathbb{C_W}$,
\method can map the two partitions $\mathbb{C_R}$ and $\mathbb{C_W}$ back to the original nodes of the graph, \ie, $\mathbb{C_P}$. 

\begin{algorithm}[!t]

        \caption{Community Initialization}
        \label{algorithm:community_initial}
        \KwIn{Original graph $G$, privacy budget $\varepsilon_1$, the number of nodes per community at first $N$
        }
        \KwOut{Preliminary community division $\mathbb{C_{P}}=\{C_1,C_2,\dots,C_{m_2}\}$}
        \textbf{Step 1: Initialization} \\
        Divide all nodes into $m_1$ communities randomly, where each community contains $N$ nodes\\
        $\mathbb{C_R} \leftarrow Randomize(G,N)$ \\
        \textbf{Step 2: Generate the weighted graph} \\
        The nodes in the same community are combined into a super-node, then the weighted graph $G_w$ 
        is obtained
        \\
        
        \textbf{Step 3: Protect the weighted graph} \label{line:weighted_graph_begin} \\
            \For{$i$ in $G_w.nodes()$}
                {
                \textbf{Step 3-1: Protect the inner weights} \\
                Obtain $\tilde{w}_{i}$ by perturbing the true inner weight $w_{i}$ \\
                $\tilde{w}_{i} \leftarrow w_{i} + Lap(\varepsilon_1,\Delta f_{i})$ \\
                
                \textbf{Step 3-2: Protect the outer weights} \\
                Obtain $\tilde{w}_{i,j}$ by perturbing the true outer weight ${w}_{i,j}$\\
                \For{$j$ in $G_w.nodes()$ }
                {
                \If{$j>i$}
                {$\tilde{w}_{i,j} \leftarrow w_{i,j} + Lap(\varepsilon_1,\Delta f_o)$}
                }
                \label{line:weighted_graph_end}
                
                }
            \textbf{Step 4: Consistency processing} \\
            $\hat{w}_{i} \leftarrow NormSub(\tilde{w}_{i}), \hat{w}_{i,j} \leftarrow NormSub(\tilde{w}_{i,j})$ \\
            $\hat{G}_w \leftarrow \hat{w}_{i},\hat{w}_{i,j}$;\\
            \textbf{Step 5: Community detection} \\
            Obtain the partitions $\mathbb{C_W}$ by adopting Louvain method \\
            $\mathbb{C_W} \leftarrow Louvain(\hat{G}_w)$ \\
        
        \textbf{Step 6: Node restoration} \\
        $\mathbb{C_P} \leftarrow \mathbb{C_R},\mathbb{C_W}$ 
       
\end{algorithm}

\mypara{Community Adjustment}
In this step, we conduct the further adjustment based on the community division results $\mathbb{C_P}$. 
Recalling the community initialization, we randomly separate $N$ nodes into a community in the beginning. 
The nodes belonging to different communities may be divided into a community, which introduces errors to the initialization partition. 
Therefore, we design \autoref{algorithm:community_adjustment} to fine-tune the community division. 
First, we use $\mathbb{C_P}$ from the community initialization as the start point of the final community division $\mathbb{C_F}$.
Then, for each node, we find its candidate community $C_h$ which has max connections to the node. 
Here, $C_h/C_e$ is a community consisting of some nodes, where $h$ or $e$ is the label of a specific community. 
$\mathbb{C_F}$ is the set of the community partitions, which includes several communities like $C_h$.
To satisfy edge-DP, we apply the exponential mechanism 
to perturb the adjustment process of the nodes. 
The sensitivity $\Delta f_c$ is 1 since the connections $k_{j,C_e}$  changes 1 at most. 
Note that adding or removing an edge will only affect two nodes in community adjustment. 
Thus, the privacy budget of each iteration is set to $0.5\varepsilon_2$. 
By iterating over the nodes and adjusting the partitions, 
we can obtain the final community division $\mathbb{C_F}$.

\begin{algorithm}[!t]

        \caption{Community Adjustment}
        \label{algorithm:community_adjustment}
        \KwIn{Original graph $G$, privacy budget $\varepsilon_2$, preliminary community division $\mathbb{C_P}=\{C_1,C_2,\dots,C_{m_2} \}$}
        \KwOut{Final community division $\mathbb{C_F}$}
    
        {
            // $m_2$ is the number of communities\\
            Initialize
            $\mathbb{C_F}=\mathbb{C_P}$ \\
            // Update $\mathbb{C_F}$ by iterating over each node \\
            \For{j in $G.nodes()$}
                {
                \textbf{Step 1: Remove the node from original community} \\
                Remove the node $j$ from original community $C_i$
                \\
                \textbf{Step 2: Choose the final community} \\
                Calculate the connections $k_{j,C_e}$ from the node $j$ to the community $C_e$ which is in $\mathbb{C_F}$ \\
                $C_h \leftarrow EM(k_{j,\mathbb{C_F}},{0.5\varepsilon_2},\Delta f_{c})$ \\
                Add the node $j$ to $C_h$
                }
    
        }
        Obtain the final community division $\mathbb{C_F}$

\end{algorithm}

\subsection{Information Extraction} 
\label{section:information_extraction}
Based on the community partition $\mathbb{C_F}$, the edges of the graph can be divided into two parts, \ie, the edges within the community and those between communities. 
If we perturb each element of the above parts, it is easy to introduce excessive noise like the existing work \cite{qin2017ldpgen}.
Since the edges of intra-community account for the majority of the entire edges, we aggregate the two types of information separately to avoid excessive noise. 
More specifically, for the edges of intra-community, \method  counts the nodes' degree sequences in their own communities, and calculates the sum of edges between community pairs for the edges of inter-community. 

As shown in~\autoref{algorithm:information_extraction}, 
in the beginning,
\method extracts the degree sequence $d$ of intra-community and the edge vector $v$ consisting of the number of edges between different communities. 
Then \method adds the Laplace noise to $d$ and $v$ separately. 
The sensitivity of degree sequence $\Delta f_d$ is 2 because the presence of an edge affects the degree of two nodes. 
The sensitivity of edge vector $\Delta f_v$ is 1. 
Note that the degree sequence and the edge vector are disjoint subsets, which construct the whole adjacency matrix together.
Therefore, these two parts can share the same privacy budget.
After the perturbation, \method conducts the consistency processing to meet the non-negative constraint. 

\begin{algorithm}[!t]

        \caption{Information Extraction}
        \label{algorithm:information_extraction}
        \KwIn{Original graph $G$, final community division $\mathbb{C_F}$, privacy budget 
        $ \varepsilon_3 $
        }
        \KwOut{Noisy degree sequence $\hat d$, noisy edge vector $\hat v$}
       
        \textbf{Step 1: Data generation} \\
        
            \For{$C_a$ in $\mathbb{C_F}$}
                {
                // Extract the degree sequence $d_a$ of community $C_a$
                 \\
                $d_a \leftarrow G$ \\
                
                }
            \For{$C_a,C_b(a \ne b)$ in $\mathbb{C_F}$}
                {
                // Extract the number of edges $v_{a,b}$ between community $C_a$ and $C_b$ \\
                $v_{a,b} \leftarrow G$}         

        \textbf{Step 2: Noise injection} \\
        // The degree sequence and the edge vector do not overlap \\
        $\tilde{d} \leftarrow d + Lap(\varepsilon_{3},\Delta f_d)$ ,
        $\tilde{v} \leftarrow v + Lap(\varepsilon_{3},\Delta f_v)$ \\
        
        \textbf{Step 3: Consistency processing} \\
        $\hat{d} \leftarrow NormSub(\tilde{d})$ ,
        $ \hat{v} \leftarrow NormSub(\tilde{v})$

\end{algorithm}

\subsection{Graph Reconstruction} 
\label{section:graph_construction}
In the paper, we choose the CL model~\cite{aiello2000random} to synthesize a graph, which can make full use of the degree information without complicated parameter settings. 
The degree distribution of the graph generated by the CL model satisfies the strictly power-law distribution~\cite{eubank2004structural}.
The graph generation consists of two parts: Intra-community edge generation and inter-community edge generation. 
Based on the degree sequence $\hat{d}$, \method calculates the connection probability $p_{u,w}$ between node $u$ and node $w$ in the same community $C$.
\begin{equation}
    p_{u \in C,w \in C} = \frac{\hat d{^{u}_{C}} \hat d{_{C}^{w}}}{\sum_{p \in C} \hat d{^{p}_{C}} }, 
    \label{eq:prob_intra_comm}
\end{equation}
where $\hat d{^{u}_{C}}$ and $\hat d{^{w}_{C}}$ represent the degrees of node $u$ and node $w$ within the community $C$,
and the denominator represents the sum of degree sequence in the community $C$. 

The edges between every two communities are encoded into a scalar before perturbation, which loses the precise endpoints information for edges. 
Thus, \method randomly rebuilds the edges between communities under the constraint of the perturbed edges' sum.
For the community $C_a$ consisting of $N_a$ nodes and the community $C_b$ consisting of $N_b$ nodes, their connections are at most $N_a \cdot N_b$.
If the perturbed edges' sum is $\hat v_{C_a,C_b}$, 
we can get the nodes' connection probability between communities as follows. 
\begin{equation}
    p_{u \in C_a, w \in C_b} = \frac{\hat v_{C_a,C_b}}{N_a N_b }
    \label{eq:prob_inter_comm}
\end{equation}

As shown in~\autoref{algorithm:graph_construction}, the synthetic graph $G_s$ is initialized as a null graph without edges. 
Then, the intra-community edge and inter-community edge are generated 
separately. 
In each community, the edges are formed by the corresponding degree sequence. 
The edges between communities are generated based on the edge vector. 
After reconstructing all edges within and between communities, we obtain the final synthetic graph $G_s$. 

\subsection{Algorithm Analysis}
\mypara{Privacy Budget Analysis}
Recalling \autoref{fig:framework}, \method has three steps, \ie, community division, information extraction, and graph reconstruction. 
The phase of community division consists of two parts, \ie, community initialization and community adjustment, which consume privacy budgets of $\varepsilon_1$ and $\varepsilon_2$ respectively.
The privacy budget consumed by the information extraction phase is $\varepsilon_3$.
The graph reconstruction phase does not touch the real data, \ie, without consuming any privacy budget. 
Therefore, the total privacy budget is $\varepsilon = \varepsilon_1+\varepsilon_2+\varepsilon_3$.
We obtain the following theorem, and the proof is deferred to \autoref{sec:proof_appendix} due to the space limitation.

\begin{theorem}
\label{throrem:PrivGraph}
\method satisfies $\varepsilon$-edge DP, where $\varepsilon=\varepsilon_1 + \varepsilon_2 + \varepsilon_3$.
\end{theorem}

\mypara{Complexity Analysis}
We compare the time complexity and the space complexity of different methods.
\hrg has the highest time complexity while \der has the 
highest space complexity.
The detailed analysis is in \autoref{sec:complexity_analysis}.

\begin{algorithm}[!t]

        \caption{Graph Reconstruction}
        \label{algorithm:graph_construction}
        \KwIn{Null graph $G_N$, final community division $\mathbb{C_F}$, noisy degree sequence $\hat d$, noisy edge vector $\hat v$}
        \KwOut{Synthetic graph $G_s$}
        Initialize $G_s$ as $G_N$ \\
        \textbf{Step 1: Intra-community edge generation} \\
        
            \For{$C_a$ in $\mathbb{C_F}$}
                {
                Generate subgraph $s_a$ based on $\hat d_a$ and~\autoref{eq:prob_intra_comm}
                 \\
                $s_a \leftarrow \hat d_a$ \\
                Update $G_s$ by $s_a$
                
                }

        \textbf{Step 2: Inter-community edge generation} \\
            \For{$C_a,C_b(a \ne b)$ in $\mathbb{C_F}$}
                {
                Generate subgraph $s_{a,b}$ based on $\hat v_{a,b}$ and~\autoref{eq:prob_inter_comm}
                \\
                $s_{a,b} \leftarrow \hat v_{a,b}$ \\
                Update $G_s$ by $s_{a,b}$
                }

\end{algorithm}

\section{Evaluation}
\label{sec:evaluation}
In~\autoref{subsec:end_to_end}, we first conduct an end-to-end experiment to illustrate the effectiveness of \method compared with the state-of-the-art methods. 
Then, we compare \method with several tailored private methods for the specific graph analysis tasks in~\autoref{subsec:comparison_tailored}.
Recalling the limitations of Louvain in~\autoref{subsec:preliminary_community_detection},
we explore the performance of \method on preserving small communities in~\autoref{subsec:preserve_small_community}.
We further demonstrate the superiority of \method on two large datasets in~\autoref{subsec:evaluation_large_dataset}.
In addition, we conduct an ablation study on the hyper-parameters of \method in~\autoref{sec:ablation_study}.
Finally, we show the real-world utility of \method through a practical case in~\autoref{subsec:case_study_IM}.

\subsection{Experimental Setup}
\label{subsec:experimental_setup}

\mypara{Datasets}
We run experiments on the six real-world datasets. 
\autoref{table:dataset_statistics} shows the basic information of the datasets, and the details of the datasets are refered to~\autoref{appendix_datasets}. 

\mypara{Metrics}
We evaluate the quality of the synthetic graph from five different aspects. 
Due to space constraints, we defer the detailed calculation formula of the metrics to~\autoref{appendix_metrics}.
\begin{itemize}[itemsep=2pt,topsep=2pt,parsep=0pt]
    \item \textbf{Community Discovery.} 
    For community detection, the metric mainly focuses on the similarity of the communities obtained from the original graph and the synthetic graph. 
    Hence, we choose Normalized Mutual Information (NMI)~\cite{kvalseth1987entropy} to measure the quality of community detection. 

    \item \textbf{Node Information.} 
    We utilize the eigenvector centrality (EVC) score to rank the nodes, which can identify the most influential nodes in a graph. 
    More specifically, we compare the percentage of common nodes in the top 1\% most influential nodes of the original graph and the synthetic graph. 
    Besides, we calculate the Mean Absolute Error (MAE) of the top 1\% most influential nodes’ EVC scores. 
    
    \item \textbf{Degree Distribution.}
    We adopt Kullback-Leibler (KL) divergence~\cite{kullback1997information} to measure the difference of the degree distributions between the original graph and the synthetic graph. 

    \item \textbf{Path Condition.} 
    The path condition reflects the connectivity between the nodes in the graph. 
    We provide the Relative Error (RE) of the diameters
    from the original graph and the synthetic graph. 

    \item \textbf{Topology Structure.} 
    We adopt clustering coefficient~\cite{wasserman1994social} and modularity~\cite{guimera2004modularity} to reflect the topology structure of the graph. 
    We compare their REs between the original graph and the synthetic graph.
    
\end{itemize}

\begin{table}[!t]
    \caption{Dataset Statistics.}
    \vspace{-0.3cm}
    \centering
    \footnotesize
    \setlength{\tabcolsep}{0.7em}
    \begin{tabular}{c | c | c | c | c}
    \toprule
     \textbf{Dataset} & \textbf{Nodes} & \textbf{Edges} & \textbf{Density} & \textbf{Type} \\
     \toprule
       Chamelon~\cite{rozemberczki2021multi}  & 2,277 & 31,421 & 0.01213 
       & Web page
       \\
       Facebook~\cite{leskovec2012learning}  & 4,039 & 88,234 & 0.01082 
       & Social 
       \\
       CA-HepPh~\cite{leskovec2007graph} & 12,008 & 118,521 & 0.00164 & Collaboration  \\
       Enron~\cite{dataset-enron} & 33,696 & 180,811 & 0.00032 & Email  \\ 
        Epinions~\cite{richardson2003trust} & 75,879 & 405,740 & 0.00014 & Trust \\
       Gowalla~\cite{cho2011friendship}
       & 196,591 & 950,327 & 0.00005 & Social \\
      \bottomrule
    \end{tabular}
    
    \label{table:dataset_statistics}
\end{table}

\mypara{Competitors} 
We compare \method with \hrg\cite{xiao2014differentially}, \der\cite{chen2014correlated},  \tmf\cite{nguyen2015differentially},
and \gen\cite{qin2017ldpgen} introduced in~\autoref{section:existing_solutions}. 
For a fair comparison, we adopt the recommended parameters from the original papers. 
Since \gen is originally designed for LDP, we reproduce the method in a DP way to ensure the rationality of comparison.
Due to the high space complexity of \der, it is hard to run it on the Enron dataset. 
We provide the result on the other three datasets. 

\mypara{Experimental Settings}
For \method, we set the number of nodes $N=20$ in community initialization, and set the resolution parameter $t=1$ in Louvain.

\mypara{Implementation}
For the allocation of privacy budget, we set $\varepsilon_1 = \varepsilon_2 = \varepsilon_3 = \frac{1}{3} \varepsilon$, 
where the total privacy budget $\varepsilon$ ranges from 0.5 to 3.5. 
We implement \method with Python 3.8, and all experiments are conducted on a server with AMD EPYC 7402@2.8GHz and 128GB memory.
We repeat experiment 10 times for each setting, and provide the mean and the standard variance. 

\subsection{End-to-End Evaluation}
\label{subsec:end_to_end}

\begin{figure*}[!t]
\centering
\includegraphics[width=0.9\textwidth]
{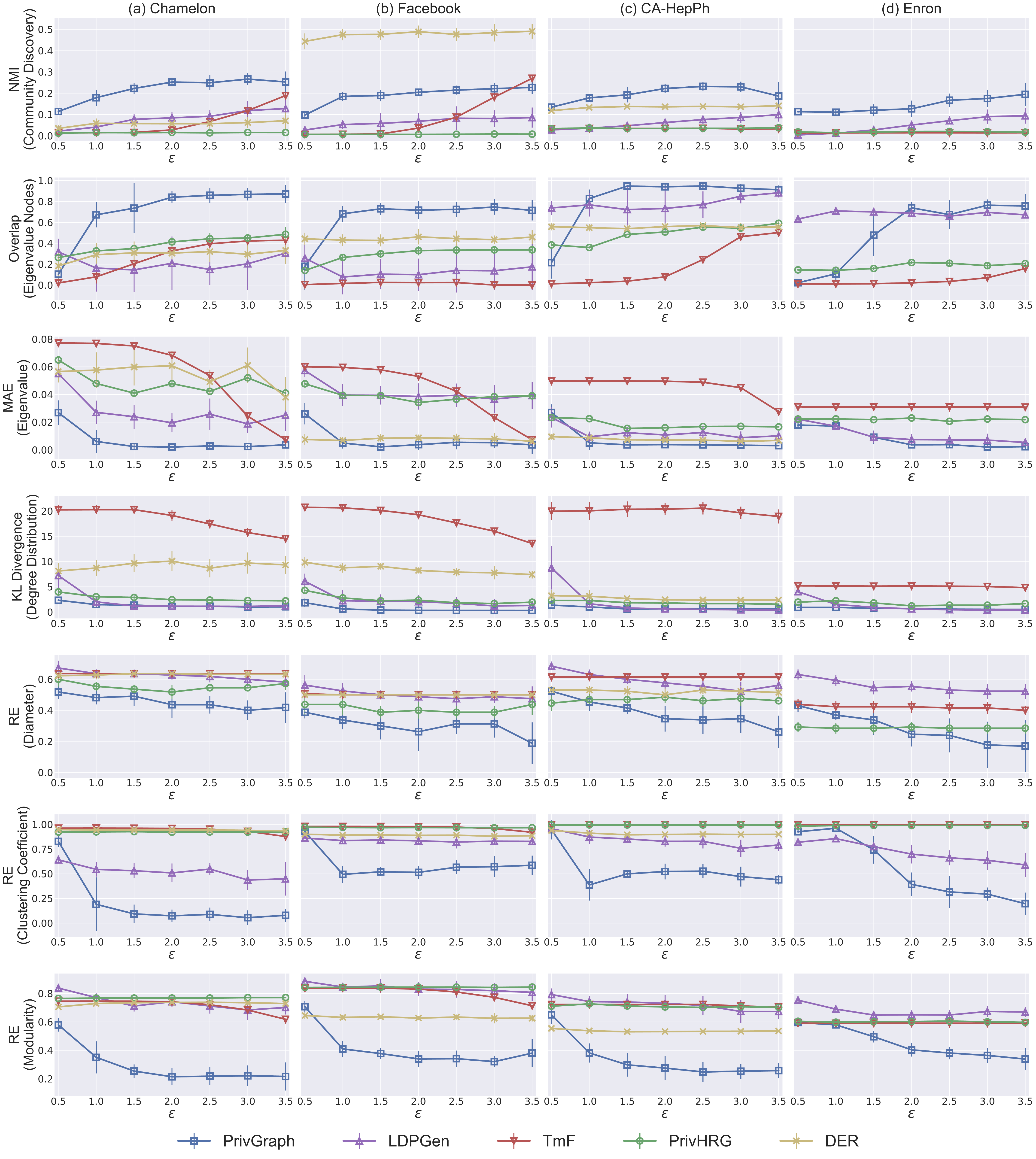}
\vspace{-0.4cm}
\caption{
{End-to-end comparison of different methods.
The columns represent the used datasets, and the rows stand for different metrics.
In each plot, the x-axis denotes the privacy budget $\varepsilon$, and the y-axis denotes the performance. For the first two rows, higher is better. For the last five rows, lower is better.}}
\label{fig:end_to_end}
\vspace{-0.3cm}
\end{figure*}

In this section, we perform an end-to-end evaluation of \method and the competitors from five perspectives. 
\autoref{fig:end_to_end} illustrates the experimental results on four datasets.

\mypara{Results on Community Discovery}
The first row of \autoref{fig:end_to_end} illustrates the NMI results on four datasets, where a higher value of NMI stands for higher accuracy. 
We have the following observations from the NMI results. 
1) \method shows similar trends for each dataset,
and the NMI values of \method tend to upward with the increase of privacy budget. 
2) \der performs better for the Facebook dataset
because it is a social network dataset, 
and the nodes are more closely clustered with each other.
\der is designed for this densely connected adjacency matrix, 
and the closely connected nodes can be easily divided into the same sub-regions.
For the other two datasets, \der
does not perform as well as \method.
3) The NMI values of \gen increase with the privacy budget, but not as obviously as \method. 
The reason is that the generated groups are not precise and do not form effective communities.
4) For \hrg, the NMI values do not improve significantly with increased privacy budgets. 
\hrg reconstructs the edges based on the probability between nodes, which lacks the consideration of community structure. 
5) The NMI values of \tmf are close to 0 under all privacy budgets on CA-HepPh and Enron datasets.
While on Chamelon and Facebook datasets, the NMI values begin to increase when $\varepsilon$ is larger than 2. 
Since \tmf directly adds Laplace noise to all elements in adjacency matrix, 
the overall perturbation strength of \tmf is related to the number of nodes and privacy budget. 
The Chamelon and Facebook datasets have less amount of nodes than the other two datasets, thus \tmf obtains higher NMI values on Chamelon and Facebook datasets with the same privacy budget. 
Thus, \tmf is not fit for protecting large graphs under strong privacy protection requirements. 

\mypara{Results on Node Information}
The second row illustrates the overlap of nodes in the top 1\% eigenvalues between the original and synthetic graphs.
We have the following observations from the results. 
1) \method behaves better than existing methods when $\varepsilon \geq 2$. 
Since the nodes within a community have similar connecting structures, we reconstruct the edges with more coarse-grained nodes' degree distributions, which reduces the perturbation noise. 
\method can effectively keep in line with the original adjacency matrix, thus enabling a higher coverage of top eigenvalues than other methods. 
When the privacy budget is small, the community division is not accurate enough, and the perturbation in the reconstruction
phase is significant, thus causing a low overlap rate.
For the first three datasets, once the privacy budget is up to 1,
\method achieves a high overlap rate. 
However, for the Enron dataset, it requires a larger privacy budget to ensure a higher recovery accuracy due to its larger size and the number of communities formed.
2) \gen performs well on the last two datasets and poorly on the first two datasets.
The reason is that the difference in the EVC scores between the influential nodes for the first two datasets is smaller, which is susceptible to noise interference.
3) \der adaptively identifies dense regions of the adjacency matrix by a data-dependent partitioning process, which aims to preserve the structure of the original graph, and thus performs better than \tmf and \hrg. 
4) We can see that \method performs better than existing methods in the MAE of most influential nodes in most cases. 
Due to the excessive noise added, the MAE obtained by other methods is larger.

\mypara{Results on Degree Distribution}
We evaluate the performance of different methods on the degree distribution of nodes.
From the fourth row of \autoref{fig:end_to_end}, we have the following observations. 
1) The KL divergence of \method is low on four datasets.
\method reconstructs the most edges of graph based on the nodes' degree within the community to balance information loss and noise injection well. 
Thus, the degree distribution of \method is similar to that of the original graph. 
2) \gen adopts the same granularity to retain information within and between groups, but the number of edges from the node to other groups is actually low.
Therefore, the KL divergence of \gen is high when the privacy budget is small because it generates much higher than the real number of edges.
For the first two datasets, the proportion of low-degree nodes is higher, and the noise is prone to corrupt the original degree distribution.
3) For \hrg, nodes with a higher degree are more likely to form connected edges. 
\hrg maintains the connection features of the nodes and obtains a prior performance than \der and \tmf. 
4) \der divides the whole adjacency matrix into lots of sub-regions. 
For the sparse sub-regions, \ie, the nodes' degrees are small, the injected noise usually overwhelms the degree of nodes. 
5) \tmf performs worst due to the random selection of the true edges,  and it does not consider the degree distribution.

\mypara{Results on Path Condition} 
The fifth row of \autoref{fig:end_to_end} illustrates the RE of the diameter of different methods on four datasets. 
1) \method acquires more accurate diameters than the competitors. 
The diameter reflects the connectivity of the graph. 
Recalling \autoref{section:overview}, \method conducts community division and extracts graph information from intra- and inter-community by different granularities. 
In this way, \method preserves the graph structure throughout the perturbation process.
For CA-HepPh and Enron datasets, when the privacy budget is small, the injected noise is high because of more nodes and communities, which induces large RE values.
2) \hrg generates a graph based on the edges between node pairs, \ie, path of the graph.
The diameter is a specific set of edges, which is closely related to the paths, thus \hrg obtains lower RE values than \der, \gen and \tmf. 
3) \der divides the whole adjacency matrix into pieces, thus it is hard for \der to reconstruct the diameter. 
4) The grouping quality of \gen is not precise, so it is challenging to recover the diameter.
5) \tmf randomly selects 1-cells in adjacency matrix without considering graph structure, which leads to higher RE values than other strategies. 

\mypara{Results on Topology Structure}
We evaluate the performance of the topology structure from the clustering coefficient aspect.
From the sixth row of \autoref{fig:end_to_end}, we have the following observations. 
1) \method outperforms other methods in most cases, which indicates that \method can better recover the clustering information of the original graph. 
2) The accuracy of the clustering coefficient is strongly related to the number of closed triplets and open triplets.
A triplet is three nodes connected by two (open triplets) or three (closed triplets) edges.
By precise community division and appropriate noise perturbation, \method improves the accuracy of the number of triplets.

Moreover, we verify the effectiveness of \method on modularity. 
As shown in the last row of~\autoref{fig:end_to_end}, we have the following observations.
1) The RE of \method is less than other methods. 
\method obtains a great partition in the first phase and extracts the information properly. 
When the privacy budget is 0.5, the RE of \der is close to \method. 
The reason is that the privacy budget is divided into multiple parts, and too small privacy budget declines the utility of community division and information extraction. 
2) The RE of \der is less than the other three methods. 
\der divides the adjacency matrix into a number of blocks, thus maintaining a certain structural information. 
3) \hrg perturbs the connection probabilities between nodes. It is too fine-grained and cannot recover the structure of graph well.
4) Since \tmf selects true edges randomly,
when the privacy budget is small, it is difficult to rebuild a structure that is similar to the original graph.  
5) The RE of \gen is high because the noise is injected at the same granularity within and between groups,
which does not preserve the original graph's structure well.

{
\mypara{Takeaways}
In general, the performance of \method is better than other methods in most cases. 
Based on the analysis and experimental results, we obtain the following conclusions. 
\begin{itemize}[itemsep=0pt,topsep=2pt,parsep=0pt]
    \item \method aggregates the similar nodes in the same community, which significantly reduces the dimension of the original data and helps to obtain an accurate community discovery and topology structure.
    \item \method extracts the information of intra-community and inter-community at various granularities and adopts different approaches to reconstruct the edges within and between communities.
    Thus, the node information, the degree distribution, and the path condition can be retained well.
    \item When the privacy budget is small, the community partition may be inaccurate because of the strong  perturbation noise, which impacts the accuracy of the final results. 
    \item Both \gen and \der perform well in the aspect of community division due to the grouping operations in their workflow. \hrg reconstructs the graph based on the probability of connection between two nodes, thus it has a good performance on the path condition. \tmf perturbs the adjacency matrix directly and requires a large privacy budget to achieve competitive results.
\end{itemize}
}

\subsection{Comparison with Tailored Methods}
\label{subsec:comparison_tailored}
In this section, we compare \method with tailored private methods on three metrics, \ie, degree distribution~\cite{wang2013preserving}, clustering coefficient~\cite{imola2021locally}, and modularity~\cite{nguyen2016detecting}, which are the widely used metrics in graph analysis~\cite{wei2020asgldp}, and there are many existing works optimized for them~\cite{hay2009accurate,wang2013preserving,ye2020towards,imola2021locally,nguyen2016detecting}.

\begin{figure*}[!t]
    \centering
    \includegraphics[width=0.85\textwidth] {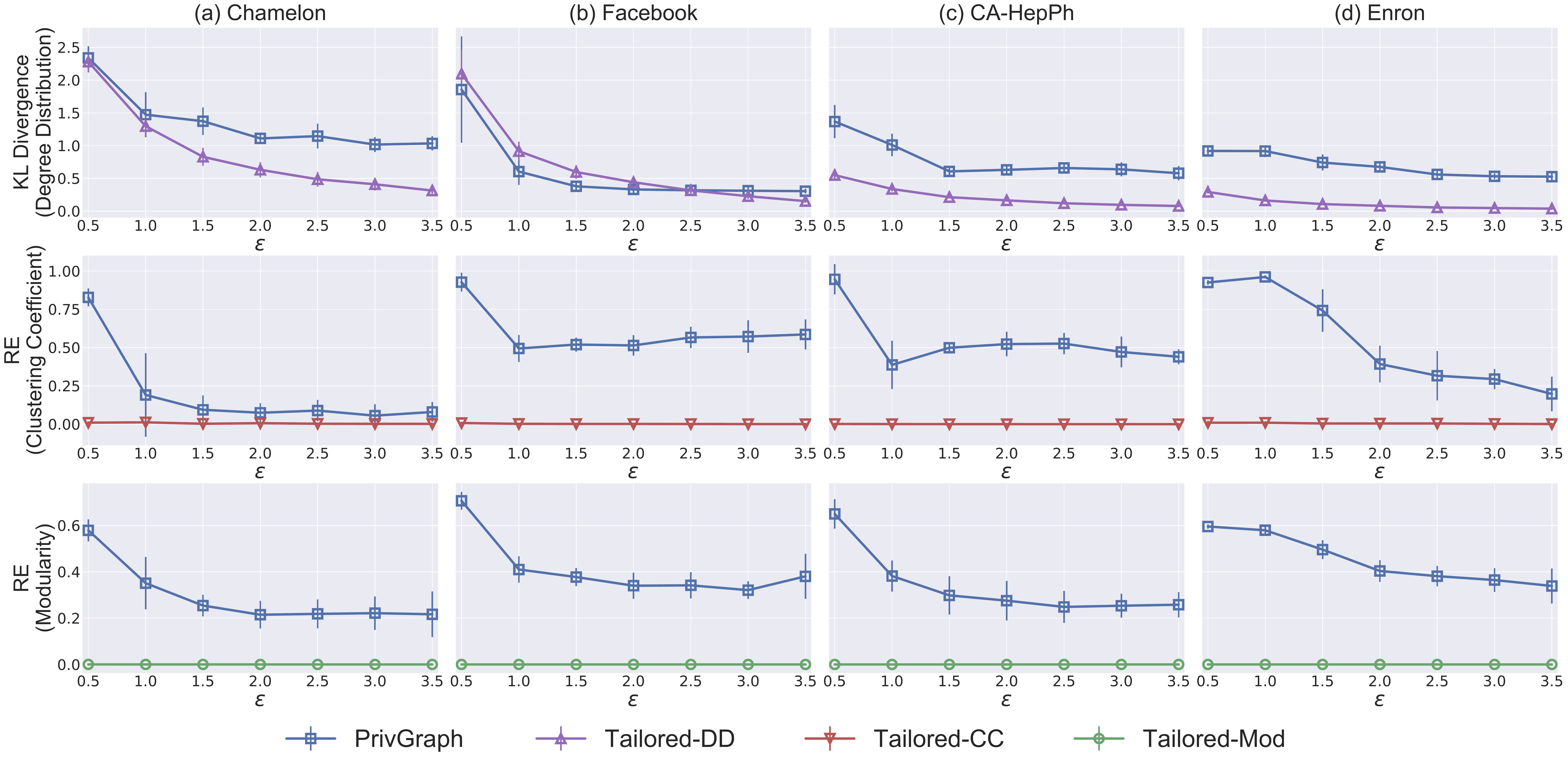}
    \vspace{-0.4cm}
    \caption{
    [Lower is better]
    Comparison with tailored methods. 
    The columns represent the used datasets and the rows stand for different metrics. 
    In each plot, the x-axis denotes the privacy budget $\varepsilon$, and the y-axis denotes the performance. 
    }
    \vspace{-0.5cm}
    \label{fig:compare_tailor}
\end{figure*}

\autoref{fig:compare_tailor} illustrates the performance 
on four datasets.
We name the tailored methods for above three metrics as Tailored-DD, Tailored-CC, and Tailored-Mod. 
In general, we observe that \method achieves competitive performance on the degree distribution, 
yet performs worse than the tailored methods on the clustering coefficient and the modularity.
For degree distribution, the performance of \method is close to Tailored-DD. 
The reason is that \method reconstructs the edges of intra-community by nodes' degree, resulting in small KL divergence.
Interestingly, the KL divergence of \method on the Facebook dataset is even smaller than Tailored-DD when the privacy budget is low.
This can be explained by the fact that the Facebook dataset is a social network, and the nodes tend to reside in more compact communities.
\method can partition the nodes into the corresponding communities precisely.
Then, the degree information of nodes within the community is extracted and reconstructed. 

For the clustering coefficient and the modularity,
the REs of Tailored-CC and Tailored-Mod are smaller than 0.012 on four datasets.
The reason is that the clustering coefficient and the modularity are published as a single value instead of a series of values like degree distribution.
For tailored methods, the information to be perturbed is highly concentrated, and the entire privacy budget is used to protect the single value. 
However, \method requires generating a whole graph, 
which is designed from more perspectives and needs to divide the privacy budget into multiple parts. 

\subsection{Preservation for Small Communities}
\label{subsec:preserve_small_community}
\begin{figure*}[!t]
    \centering
    \includegraphics[width=0.9\textwidth] {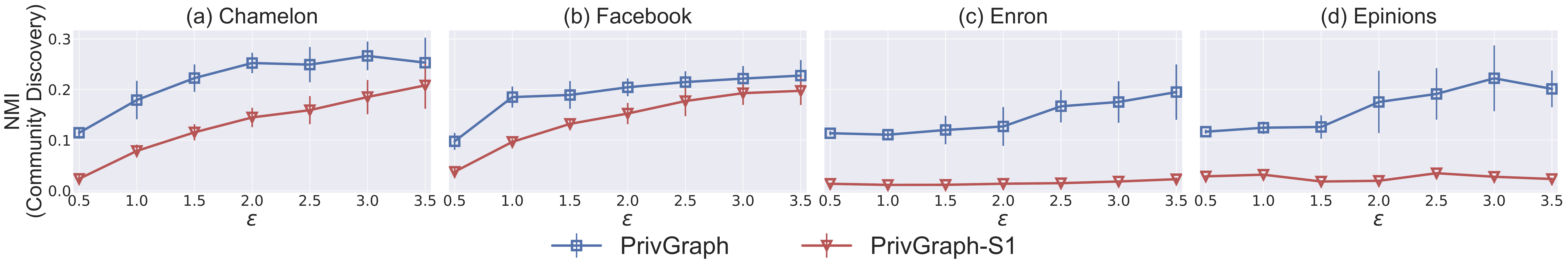}
    \vspace{-0.4cm}
    \caption{
    [Higher is better]
    Comparison of \method and \method-S1. 
    The columns represent the used datasets and the rows stand for different metrics. 
    In each plot, the x-axis denotes the privacy budget $\varepsilon$, and the y-axis denotes the NMI value. 
    }
    \vspace{-0.5cm}
    \label{fig:compare_nmi_S1}
\end{figure*}

Recalling \autoref{subsec:preliminary_community_detection}, \method's information extraction and graph reconstruction can compensate for Louvain's limitations, specifically in cases where the modularity optimization may result in small communities being overlooked. 
We compare the accuracy of the community division between the first phase of \method (called \method-S1) and the whole processes of \method. 
More specifically, leveraging the community division results from the original graph as the baseline, we use NMI to measure the similarity of the results from \method-S1 and \method to the original graph. 
For a fair comparison, \method-S1 exhausts the entire privacy budget that is distributed among three components in \method. 
\autoref{fig:compare_nmi_S1} illustrates the comparison results. 

\method performs better than \method-S1 since the information extraction and graph reconstruction processes help to recover the small communities lost in Louvain. 
In addition, compared to \method, \method-S1 shows different tendencies in the first two datasets and the last two datasets, where Enron and Epinions have a greater number of small communities compared to Chamelon and Facebook. 
Since \method-S1 has the potential to incorrectly merge small communities into larger ones, resulting in low NMI values for the last two datasets. 
For \method, the phases of information extraction and graph reconstruction are beneficial to retain small communities.
Although small communities are merged into large ones in the first phase, the tightly connected edges still exist in the same community, \ie, the information of small communities is preserved in the degree distribution. 
Then, the degree distribution of intra-community is extracted and reconstructed in the final phase, which contributes to the restoration of small communities.
Therefore, \method achieves great performance on all four datasets. 

\section{Related Work}
\label{sec:related_work}

\subsection{Differentially Private Graph Analysis}

Differentially private graph analysis can accomplish a series of statistical tasks on private data.
Several strategies are designed for various downstream tasks, including degree distribution, community division, clustering coefficient, \etc

\mypara{Degree Distribution}
There are some works dedicated to the task of degree release~\cite{kasiviswanathan2013analyzing,raskhodnikova2016lipschitz,hay2009accurate}.
In particular, 
lipschitz extensions and exponential mechanism are used in \cite{raskhodnikova2016lipschitz} for approximating the degree distribution of a sensitive graph.
Hay \etal~\cite{hay2009accurate} designed an algorithm for publishing the degree distribution based on a constrained inference technique.
 
\mypara{Topology Structure}
Many studies have investigated the problem of protecting the topology structure of a graph~\cite{nguyen2016detecting,ji2020community,chen2020publishing,zhang2020community,jorgensen2016publishing}, yet they are fundamentally different from \method.
First, for the problem definition, prior works
~\cite{chen2020publishing,zhang2020community,jorgensen2016publishing,ji2020community} require both node attributes and edges, while \method only can touch the edges.
Second, the results are published at different granularities. 
The related works~\cite{ji2020community,nguyen2016detecting} output the community partitions, \ie,  coarse granularity. \method generates a complete graph, \ie, fine granularity.
Nguyen \etal~\cite{nguyen2016detecting} proposed to form a weighted graph by a filtering technique and then apply Louvain method to detect the communities.
However, the random grouping of nodes in the beginning may cause a large deviation. 
Ji \etal~\cite{ji2020community} formulated the community detection of attributed graph as a maximum log-likelihood problem.

\subsection{Privacy Attacks on Graph}
There exist several works related to privacy attacks on graph data~\cite{Zhang2020TowardsPG,xian2021towards,zhang2022inference,yang2017bi,eagle2009inferring,SHZCYBZS22}.
According to different assumptions of the attacker's ability, the inference attacks on edges could be divided into two categories. 
The first type of attacks make inferences based on the edge structure information~\cite{Zhang2020TowardsPG,xian2021towards},
\ie, nodes with higher structural similarity tend to be connected to each other.
The second type of attacks attempt to reconstruct the original graph mainly based on the node features~\cite{yang2017bi,eagle2009inferring},
\ie, nodes with more similar attributes tend to link more closely.

For the first type of inference attacks on edges, \method could defend the attacks effectively since they do not violate edge-DP guarantees.
For the second type of attacks, a single edge-DP technique may prove insufficient to thwart them 
since the attacker can infer edges barely based on the node features. 
In addition, it should be noticed that the ideal application scenarios of edge-DP require that all edges are independent of each other. 
In practice, different edges may be correlated, \ie, the presence of an edge might be inferred by other edges, which 
introduces greater challenges to privacy protection. 
The problem could be mitigated by applying $k$-edge DP~\cite{hay2009accurate} and considering the correlation between edges, 
which are promising directions for further improvements. 

There are some de-anonymization attacks~\cite{lee2018quantify,zhao2020using} to re-identify nodes in an anonymized graph according to an auxiliary graph and a set of seed mappings, which is orthogonal to the privacy guarantee provided by edge-DP~\cite{hay2009accurate,dwork2006calibrating}.

\subsection{Differentially Private Data Synthesis}

There are some existing studies for other types of data.

\mypara{Tabular Data}
There are three mainstream methods to process tabular data: graphical model-based, game-based, and deep generative model-based.
The core idea of graphical model-based methods is to estimate a graphical model which approximates the distribution of the original dataset under DP~\cite{mckenna2019graphical,zhang2021privsyn}. 
The game-based methods regard the dataset synthesis problem as a zero-sum game~\cite{gaboardi2014dual,vietri2020new}. 
Deep generative model-based methods first train a deep generative model under DP and adopt the model to synthesize the dataset~\cite{beaulieu2019privacy,frigerio2019differentially}. 

\mypara{Trajectory Data}
There are a few works that investigate the synthesis of trajectory dataset while satisfying DP~\cite{he2015dpt,gursoy2018utility,chen2012n-grams,wang2023privtrace}.
He \etal~\cite{he2015dpt} designed DPT method to discretize the space by various granularities and built multiple prefix trees.
Wang \etal~\cite{wang2023privtrace} proposed to dynamically choose between first-order and second-order Markov models to tradeoff noise error and correlation error.
Du \etal~\cite{DHZFCZG23} proposed LDPTrace for the local DP settings. 

\section{Conclusion}
\label{sec:conclusion}
In this paper, we propose \method for publishing graph data under DP.
By exploiting community information and considering the different characteristics of connections within and between communities, we extract the structure of the graph effectively and reconstruct it accurately. 
\method satisfies rigorous DP while achieving a balance between information loss and perturbation strength.
Extensive experiments on six real-world datasets demonstrate the superiority of \method. 
Compared with tailored private methods optimized for specific graph analysis tasks, \method still shows competitive results on some settings.
We also explore the performance of \method on preserving small communities.
Then, we empirically analyze the impact of hyper-parameters of \method. 
For the practical applications, we show the advantage of \method on the influence maximization problem. 

\section*{Acknowledgments} 
We thank the anonymous shepherd and reviewers for their constructive feedback.
This work was supported in part by the National Natural Science Foundation of China under Grants 62103371, U20A20159, the Helmholtz Association within the project ``Trustworthy Federated Data Analytics'' (TFDA) (No. ZT-I-OO1 4), and CISPA-Stanford Center for Cybersecurity (FKZ:13N1S0762).

{
    \footnotesize
    \bibliographystyle{abbrv}
    \bibliography{easy}
}

\appendix

\section{Proof of \autoref{throrem:PrivGraph}}
\label{sec:proof_appendix}

\method consists of three phases: Community division, information extraction, and graph reconstruction.
Specifically, the community division phase includes two parts: \ie community initialization and community adjustment.
Next,
we show that the components of \method are satisfying edge-DP.

\smallskip
\mypara{Proof 1: Community Initialization satisfies $\varepsilon_1$-edge DP}
\begin{proof}
\label{proof:community_initial_DP}
In the process of community initialization, \method will perturb the true weighted graph generated from the random partitions.
The perturbations of inner weights and outer weights are achieved by adding Laplace noise.
Recalling \autoref{subsec:differential privacy}, Laplace Mechanism can provide rigorous differential privacy guarantee.
The inner and outer weights are independent of each other.
According to the parallel composition, they can share the same privacy budget, \ie, $\varepsilon_1$.
The consistency processing and community detection do not touch the true data and do not consume privacy budget.
Hence, community initialization satisfies $\varepsilon_1$-edge DP.
\end{proof}

\mypara{Proof 2: Community Adjustment satisfies $\varepsilon_2$-edge DP}

\begin{proof}
\label{proof:community_adjust_DP}
\method adopts exponential mechanism to select the community for each node with the privacy budget of $\varepsilon_a=0.5\varepsilon_2$.
We can assume that the two nodes corresponding to the only edge that differs between the original graph and the edge neighbor graph are $i$ and $j$.

For the nodes $i$ and $j$, given any inputs $v_1,v_2$ ($v_1$ and $v_2$ differ by an edge), and output $o$, combining the probability equation of~\autoref{subsec:differential privacy}, we have

\begin{equation*}
    \begin{aligned}
    \frac{\Pr {\mathcal{A}_q(v_1)=o }}{\Pr {\mathcal{A}_q(v_2) = o} } 
    & \le
    e^{\varepsilon_a},
    \end{aligned}
\end{equation*}
where $q$ is the quality function, $\mathcal{O}$ is the set of all possible outputs, and $\Delta f_c$ is the global sensitivity.

For the other nodes except for node $i$ and node $j$, 
the inputs $v_1$ and $v_2$ are the same.
Therefore, we have
\begin{equation*}
    \frac{\Pr {\mathcal{A}_q(v_1)=o }}{\Pr {\mathcal{A}_q(v_2) = o} }
    = e^{0}. 
\end{equation*}

Based on the sequential composition,
we can obtain $0.5\varepsilon_2+0.5\varepsilon_2+0=\varepsilon_2$.
Hence, community adjustment satisfies $\varepsilon_2$-edge DP.
\end{proof}

\mypara{Proof 3: Information Extraction satisfies $\varepsilon_3$-edge DP}

\begin{proof}
\label{proof:information_extraction_DP}
In the phase of information extraction, the original degree sequence of intra-community and the true edge vector between communities are injected Laplace noise, respectively. 
Note that the degree sequence and the edge vector are disjoint subsets,
they can be perturbed by the same privacy budget.
The proof is similar to community initialization.
Based on the parallel composition,
we can obtain that information extraction satisfies $\varepsilon_3$-edge DP.
\end{proof}

\mypara{Overall Privacy Budget}
According to the above proofs,
in the first phase,
community initialization satisfies $\varepsilon_1$-edge DP and community adjustment satisfies $\varepsilon_2$-edge DP.
The phase of information extraction satisfies $\varepsilon_3$-edge DP.
In graph reconstruction, \method processes the perturbed data without consuming privacy budget.
Hence, \method satisfies $\varepsilon$-edge DP in accordance with \textit{sequential composition}.

\section{Complexity Analysis}
\label{sec:complexity_analysis}

In this section, we analyze the computational complexity of various methods, and quantitatively evaluate their running time and memory consumption. 

\mypara{Time Complexity}
We provide the time complexity by analyzing each phase of the algorithms. 
The number of edges is $m$, and the number of nodes is $n$.

For \method, the first phase is to divide the nodes into a number of communities by community initialization and community adjustment.
The time complexity of community initialization is $\mathcal{O}(k_{1}^{2})$, where $k_1$ is the number of super-nodes.
In the community adjustment, the time complexity is $\mathcal{O}(n)$.
In information extraction, \method separately perturbs the original degree sequence of each community and the edge vector between communities. 
And the time cost of reconstruction is similar to that of information extraction.
The time complexity of these two phases is $\mathcal{O}(k_{2}^{2})$, where $k_2$ is the number of communities.
Above all, we have $\mathcal{O}(k_{1}^{2})+\mathcal{O}(n)+\mathcal{O}(k_{2}^{2})=\mathcal{O}(n+k_{1}^{2}+k_{2}^{2}) < \mathcal{O}(n^2)$, thus the total time complexity of \method is $\mathcal{O}(n^2)$.
In fact, the number of communities is much less than $n$, so the computation time is short.

\tmf directly processes at most $2m$ 1-cells and 0-cells in the adjacency matrix. 
Therefore, the time complexity is $O(m)$, which increases with the number of edges linearly. 

For \hrg, it consists of three steps: HRG sampling, probability value perturbation, and graph generation.
The first step is finding a suitable HRG by MCMC sampling, with the time complexity of $\mathcal{O}(n^2\log{n})$.
The second step is adding Laplace noise to the probability values in the HRG and the time complexity is $\mathcal{O}(n)$.
The final step is reconstructing the graph based on the probability between nodes.
The time complexity is $\mathcal{O}(n^2)$.
$\mathcal{O}(n^2\log{n})+\mathcal{O}(n)+\mathcal{O}(n^2)=\mathcal{O}(n^2\log{n})$. 

\gen needs to group all nodes by $k$-means clustering, and to add noise to the formed matrix.
The time complexity of $k$-means clustering is $\mathcal{O}(nkt)$, where $k$ is the number of clusters and $t$ is the number of iterations.
The time complexity of adding noise is $\mathcal{O}(nk)$.
Therefore, we can obtain $\mathcal{O}(nkt)+\mathcal{O}(nk)\approx \mathcal{O}(n^2)$.
The time complexity of \gen is $\mathcal{O}(n^2)$.

\der includes three parts: Node relabeling, dense region exploration, and edge reconstruction.
In the first part, \der generates $\frac{n}{2}$ candidate swaps for relabeling and each swap involves exactly two columns and two rows. 
Therefore, the time complexity of node relabeling is $\mathcal{O}(n^2)$.
In the procedure of dense region exploration, the complexity is determined by applying EM to select the splitting points.
The time complexity of this procedure is $\mathcal{O}(n^2)$.
The third step is to reconstruct all leaving regions.
The time complexity is $\mathcal{O}(n^2)$.
Hence, the overall time complexity of \der is $\mathcal{O}(n^2)$.

\mypara{Space Complexity}
For \method, it requires storing the information of edges and super-nodes.
Therefore, the space complexity 
is $\mathcal{O}(m+n)$.
The memory consumption of \tmf is only related to the number of edges, \ie, $\mathcal{O}(m)$. 
For \hrg, it needs to get the information of edges and maintain an HRG, requiring $\mathcal{O}(m+n)$ storage.
The space complexity of \gen is $\mathcal{O}(n^2)$ because it must store the connection information of all nodes to each cluster.
The space complexity of \der is $\mathcal{O}(n^2)$ due to the count summary matrix.

\mypara{Empirical Evaluation}
\autoref{table:comparsion_running_time} and \autoref{table:comparsion_memory_consumption} show  the
running time and the memory consumption for all methods on the six datasets (see their details in \autoref{table:dataset_statistics}).
The empirical running time in \autoref{table:comparsion_running_time}
illustrates that the performance of \tmf is best because it processes the cells in a linear time without further operations.
The running time of \method and \gen are longer than \tmf since they require grouping.
\hrg and \der take much more time than \tmf, \gen and \method.
\hrg consumes huge time to sample an HRG, and \der spends lots of time dividing the adjacency matrix into small pieces.

\autoref{table:comparsion_memory_consumption} shows the memory consumption.
\der is the highest.
The reason is that \der needs to maintain a count matrix during the data processing. 
\gen is also high because it requires to storage the connection matrix of all nodes.
The memory consumption of other methods is close because their space complexities are linear to $m$ or $n$. 

\begin{table}[!t]
\caption{Comparison of computational complexity.}
    \centering
    \footnotesize
    \setlength{\tabcolsep}{0.9em}
    \begin{tabular}{c | c | c }
    \toprule
    \textbf{Methods} & \textbf{Time Complexity} & \textbf{Space Complexity}  \\
    \toprule
       \tmf  & $\mathcal{O}(m)$ & $\mathcal{O}(m)$   \\
       \hrg  & $\mathcal{O}(n^2\log{n})$  & $\mathcal{O}(m+n)$  \\
       \gen & $\mathcal{O}(n^2)$ & $\mathcal{O}(n^2)$ \\
       \der & $\mathcal{O}(n^2)$ & $\mathcal{O}(n^2)$ \\
       \method & $\mathcal{O}(n^2)$ & $\mathcal{O}(m+n)$ \\
    \bottomrule
    \end{tabular}
    
    \label{table:comparsion_computation_complexity}
\end{table}

\begin{table}[!t]
\caption{Comparison of running time (measured by seconds).}
    \centering
    \footnotesize
    \setlength{\tabcolsep}{0.5em}
    \begin{tabular}{c | c | c | c | c | c}
    \toprule
    & \multicolumn{5}{c}{\textbf{Methods}} \\
    {\textbf{Datasets}} & {\tmf} & {\method} & {\gen} & {\der} & {\hrg} \\
     \toprule
       Chamelon  & 0.22s & 1.47s  & 2.16s & 136.41s &  273.13s \\
       Facebook  & 0.68s & 4.13s  & 4.37s & 383.81s & 1580.24s \\
       CA-HepPh & 4.72s & 17.84s & 22.46s & 2750.58s &  4593.21s \\
       Enron & 31.20s & 118.05s  & 61.41s & N/A  & 33677.63s\\
       Epinions & 105.46s & 503.48s & 312.84s & N/A & N/A \\
       Gowalla & 242.75s & 2358.32s & N/A & N/A & N/A \\
      \bottomrule
    \end{tabular}
    
    \label{table:comparsion_running_time}
\end{table}

\begin{table}[!t]
\caption{Comparison of memory consumption (measured by Megabytes).}
    \centering
    \footnotesize
    \setlength{\tabcolsep}{0.5em}
    \begin{tabular}{c | c | c | c | c | c}
    \toprule
    & \multicolumn{5}{c}{\textbf{Methods}} \\
    {\textbf{Datasets}} & {\tmf} & {\method} & {\gen} & {\der} & {\hrg} \\
     \toprule
       Chamelon  & 30.88 & 36.46  & 61.52 & 151.74 &  50.93 \\
       Facebook  & 74.46 & 81.35  & 110.45 & 451.50 & 95.94 \\
       CA-HepPh & 141.31 & 240.85 & 632.97 & 3139.57 &  546.06 \\
       Enron & 210.18 & 913.13  & 4191.59 & N/A  & 3586.20\\
       Epinions & 482.67 & 2563.14 & 14210.52 & N/A & N/A \\
       Gowalla & 984.23 & 6811.05 & N/A & N/A & N/A \\
      \bottomrule
    \end{tabular}
    
    \label{table:comparsion_memory_consumption}
\end{table}

\section{Experimental Setup}

\subsection{Datasets}
\label{appendix_datasets}
The details of six datasets are as follows.

\begin{itemize}
[itemsep=2pt,topsep=2pt,parsep=0pt]
\item \textbf{Chamelon~\cite{rozemberczki2021multi}.} 
Collected from the English Wikipedia on the chamelon topic, 
it contains 2,277 nodes (\ie, articles) and 31,421 edges (\ie, mutual links).
\item \textbf{Facebook~\cite{leskovec2012learning}.} 
This dataset is collected from survey participants using a Facebook app. 
It contains 4,039 nodes (\ie, users) and 88,234 edges (\ie, connections).
\item \textbf{CA-HepPh~\cite{leskovec2007graph}.}
CA-HepPh is from the e-print arXiv covering scientific collaborations between authors' papers, which contains 12,008 nodes and 118,521 edges. 
The nodes represent the authors and the edges stand for the collaboration relationship. 
\item \textbf{Enron~\cite{dataset-enron}.}
The dataset is an email graph. 
Nodes represent the email accounts in Enron and edges represent the communications. 
It contains 33,696 nodes and 180,811 edges.

\item \textbf{Epinions~\cite{richardson2003trust}.}
The dataset is a trust network collected from a consumer review site.
It contains 75,879 nodes (\ie, users) and 405,740 edges (\ie, trust relationships).

\item \textbf{Gowalla~\cite{cho2011friendship}.}
Gowalla is from a location-based social networking website, which contains 196,591 nodes (\ie, users) and 950,327 edges (\ie, friendships).

\end{itemize}

\subsection{Evaluation Metrics}
\label{appendix_metrics}

We evaluate the quality of the synthetic graph from five different aspects. 

\begin{itemize}
[itemsep=2pt,topsep=2pt,parsep=0pt]
    \item \textbf{Community Discovery.} 
    Here, we choose Normalized Mutual Information (NMI)~\cite{kvalseth1987entropy} to measure the quality of community division. 
    In particular, we apply the Louvain~\cite{blondel2008fast} method to acquire the partitions from the original graph and synthetic graph. 
    Then we measure the difference between the group partitions by NMI. 
    
    Given two partitions $A=\{ A_1,A_2,\cdots A_R\}$ and $B=\{ B_1,B_2,\cdots B_S \}$ of a graph $G=(V, E)$, 
    the overlap between $A$ and $B$ can be represented through a contingency table $H$, 
    where $H_{ij}$ stands for the number of nodes that belong to $A_i$ and $B_j$. 
    Let $x_i$ (resp. $y_j$) denotes the sum of all elements in the $i$-th row (resp. $j$-th column) of the contingency table.
    We can obtain the NMI value between partitions $A$ and $B$ as follows. 
    \begin{equation*}
        NMI(A,B) = \frac{-2\sum_{i=1}^{R} \sum_{j=1}^{S}H_{ij}  \log\left(\frac{H_{ij}n}{x_i y_j}\right) }{\sum_{i=1}^{R} x_i\log\left(\frac{x_i}{n}\right) + \sum_{j=1}^{S}y_j\log\left(\frac{y_j}{n}\right)}
    \end{equation*}

    \item \textbf{Node Information.} 
    The eigenvector centrality (EVC) score is used to rank the nodes, which can identify the most influential nodes in a graph. 
    We compare the percentage of common nodes in the top 1\% most influential nodes of the original graph and the synthetic graph. 
    \begin{equation*}
        {Overlap}_{Node} = \frac{\left | N \cap \hat{N} \right |}{\left | N \right |},
    \end{equation*}
    where $N$ and $\hat N$ stand for the node sets of the top 1\% eigenvalues in the original graph and the synthetic graph. 
    Besides, the Mean Absolute Error (MAE) of the top 1\% most influential nodes’ EVC scores can be calculated as follows.
    \begin{equation*}
        MAE_{EVC} = \frac{1}{T}\sum_{i=1}^{T}\left | \hat{s}_i - s_i \right |,
    \end{equation*}
    where $s_i$ and $\hat{s}_i$ are the EVC scores of the $i$-th most influential node in the original graph and the synthetic graph.
    \item \textbf{Degree Distribution.}
    The Kullback-Leibler (KL) divergence~\cite{kullback1997information} is adopted to measure the difference of the degree distributions between the original graph and the synthetic graph. 
    \begin{equation*}
        D_{KL}\left(P \parallel \hat{P}\right) = \sum_{x\in \mathcal{X}}P(x)\log
        \left(\frac{P(x)}{\hat{P}(x)}\right),
    \end{equation*}
    where $P(x)$ and $\hat{P}(x)$ stand for the degree distributions of the original graph and the synthetic graph separately. 

    \item \textbf{Path Condition.} 
    We provide the Relative Error (RE) of the diameters, \ie, the maximum distance among all path-connected pairs of nodes, from the original graph and the synthetic graph. 
    \begin{equation*}
        {RE}_{Diam} =  \frac{\left | \hat{D} - D  \right |}{\max(\delta,D)},
    \end{equation*}
    where ${D}$ and $\hat{D}$ are the diameters of the original graph and the synthetic graph, respectively, and $\delta$ is a small constant to avoid a zero denominator. 

    \item \textbf{Topology Structure.} 
    We compare the RE of the clustering coefficients and the modularities between the original graph and the synthetic graph.
    The RE of modularity can be obtained by:
    \begin{equation*}
        {RE}_{Mod} = \frac{\left | \hat{Q} - Q  \right |}{\max(\delta,Q)}, 
    \end{equation*}
    where ${Q}$ and $\hat{Q}$ are the modularities of the original graph and the synthetic graph, respectively, 
    and $\delta$ is a small constant to avoid a zero denominator. 
    The RE of clustering coefficient can be calculated in a similar way.
    \begin{equation*}
        {RE}_{CC} = \frac{\left | \hat{Y} - Y  \right |}{\max(\delta,Y)}, 
    \end{equation*}
    where ${Y}$ and $\hat{Y}$ are the clustering coefficients of the original graph and the synthetic graph, respectively, 
    and $\delta$ is a small constant to avoid a zero denominator. 

\end{itemize}

\section{Evaluation on Large Datasets}
\label{subsec:evaluation_large_dataset}

\begin{figure*}[!t]
    \centering
    \includegraphics[width=0.9\textwidth] {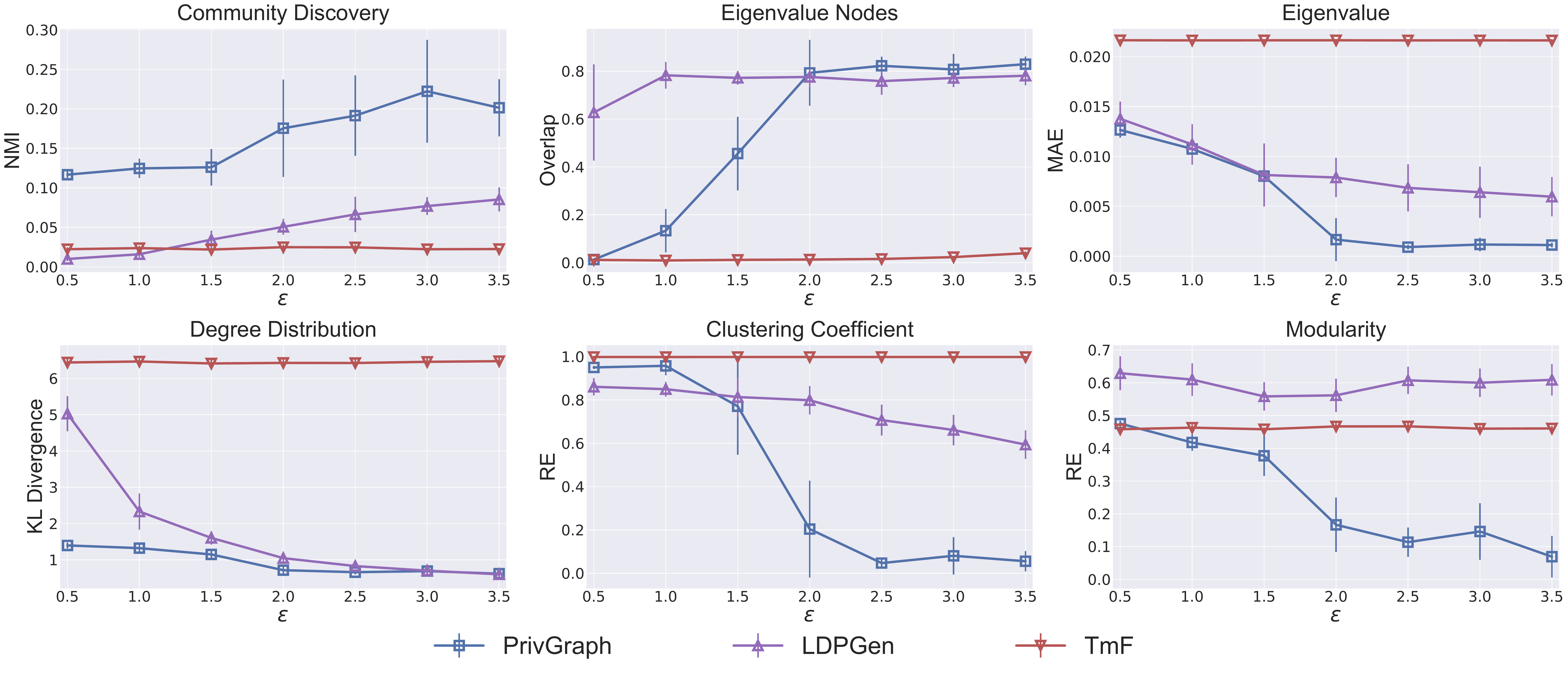}
    \vspace{-0.4cm}
    \caption{
    Performance of six metrics on the Epinions dataset. 
    In each plot, the x-axis denotes the privacy budget $\varepsilon$, and the y-axis denotes the performance. 
    For the first two metrics of the first row, higher is better. 
    For the other metrics, lower is better.
    }
    \vspace{-0.2cm}
    \label{fig:large_epinions}
\end{figure*}

\begin{figure*}[!t]
    \centering
    \includegraphics[width=0.9\textwidth] {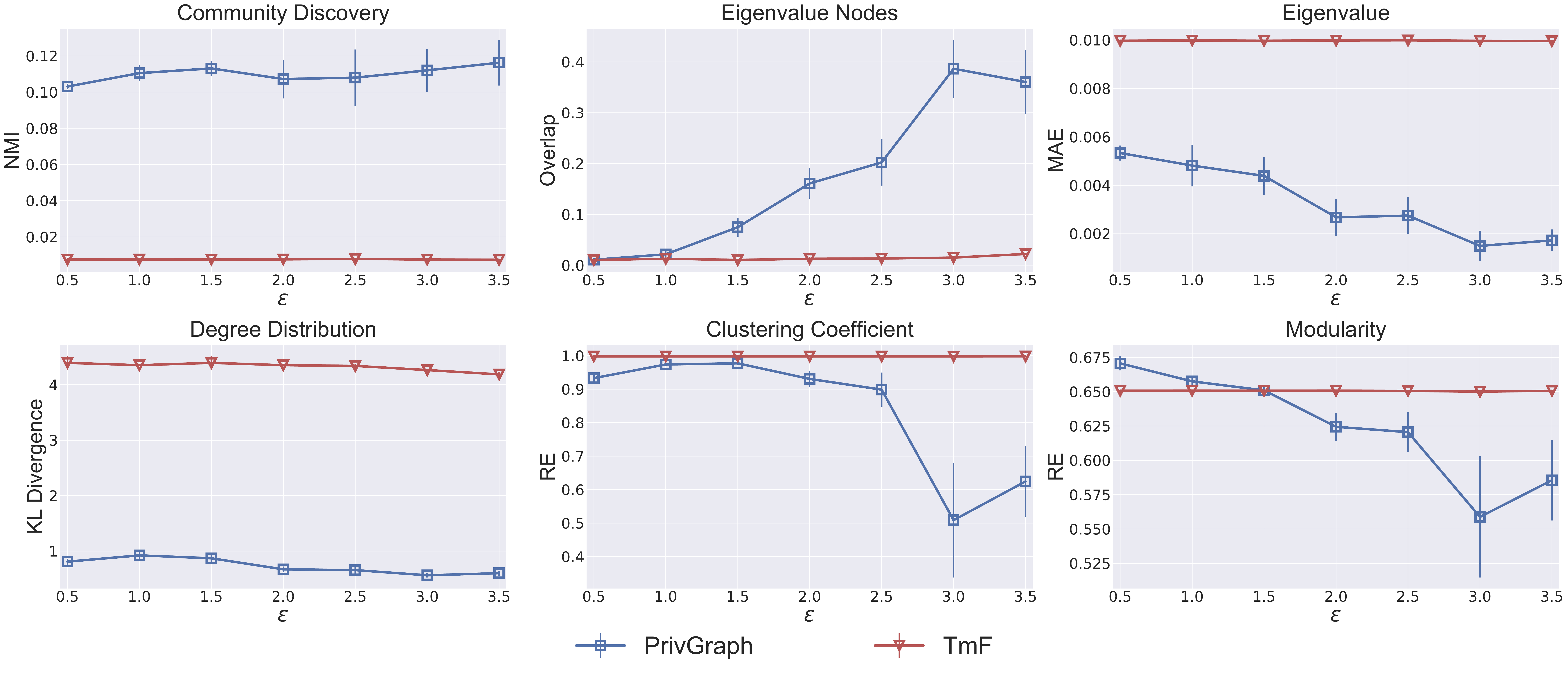}
    \vspace{-0.4cm}
    \caption{
    Performance of six metrics on the Gowalla dataset. 
    In each plot, the x-axis denotes the privacy budget $\varepsilon$, and the y-axis denotes the performance. 
    For the first two metrics of the first row, higher is better. 
    For the other metrics, lower is better.
    }
    \vspace{-0.2cm}
    \label{fig:large_gowalla}
\end{figure*}

This section includes an additional evaluation of \method and its competitors using two large datasets. 
However, due to time and space constraints, it was difficult to run \der, \hrg, and \gen on these datasets. 
Therefore, we compared \method with \tmf and \gen on the Epinions dataset, and compared \method with \tmf on the Gowalla dataset. 
The experimental results are shown in \autoref{fig:large_epinions} and \autoref{fig:large_gowalla}, while the diameter metric is not included due to the lengthy calculation time.

\method continues to exhibit significant advantages over other methods when dealing with large datasets. 
This is due to its ability to group similar nodes into the same community, thus preserving the original graph structure. 
Additionally, \method is able to extract both intra- and inter-community information at different levels of granularity, 
and use appropriate reconstruction methods. 
As a result, \method generally outperforms other methods in most scenarios. 
However, when the privacy budget is limited, the performance of certain metrics may be impacted due to the strong perturbation. 
In contrast, \gen achieves superior results compared to \tmf by reducing the dimensionality of the original graph through clustering. 
\tmf, on the other hand, perturbs the entire adjacency matrix, and requires a large privacy budget to achieve competitive results. 

\section{Ablation Study}
\label{sec:ablation_study}

\subsection{Impact of the Number of Nodes for Community Initialization}
\label{subsec:impact_iteration} 
Recalling \autoref{section:community_division}, the number of nodes $N$ in community initialization controls the initial scale of the communities. 
If $N$ is too small, 
there will generate a large number of communities, and the perturbation noise tends to overwhelm the true information, especially when the privacy budget is small.
If $N$ is too high, 
it will aggravate the error caused by random division and cannot obtain a precise partition.
Therefore, a suitable $N$ value is significant to balance the above two factors. 
However, the theoretical selection of $N$ is hard due to the following two reasons. 
Firstly, the graph data has various structures in practice, \eg, the number of communities is usually different in graph datasets. 
In addition, \method can access the original graph by using privacy budget, 
yet the complexity of estimating an accurate $N$ brings a high privacy budget overhead, which is not cost-effective for other processes of \method. 

\begin{figure*}[!t]
    \centering
    \includegraphics[width=0.95\textwidth] {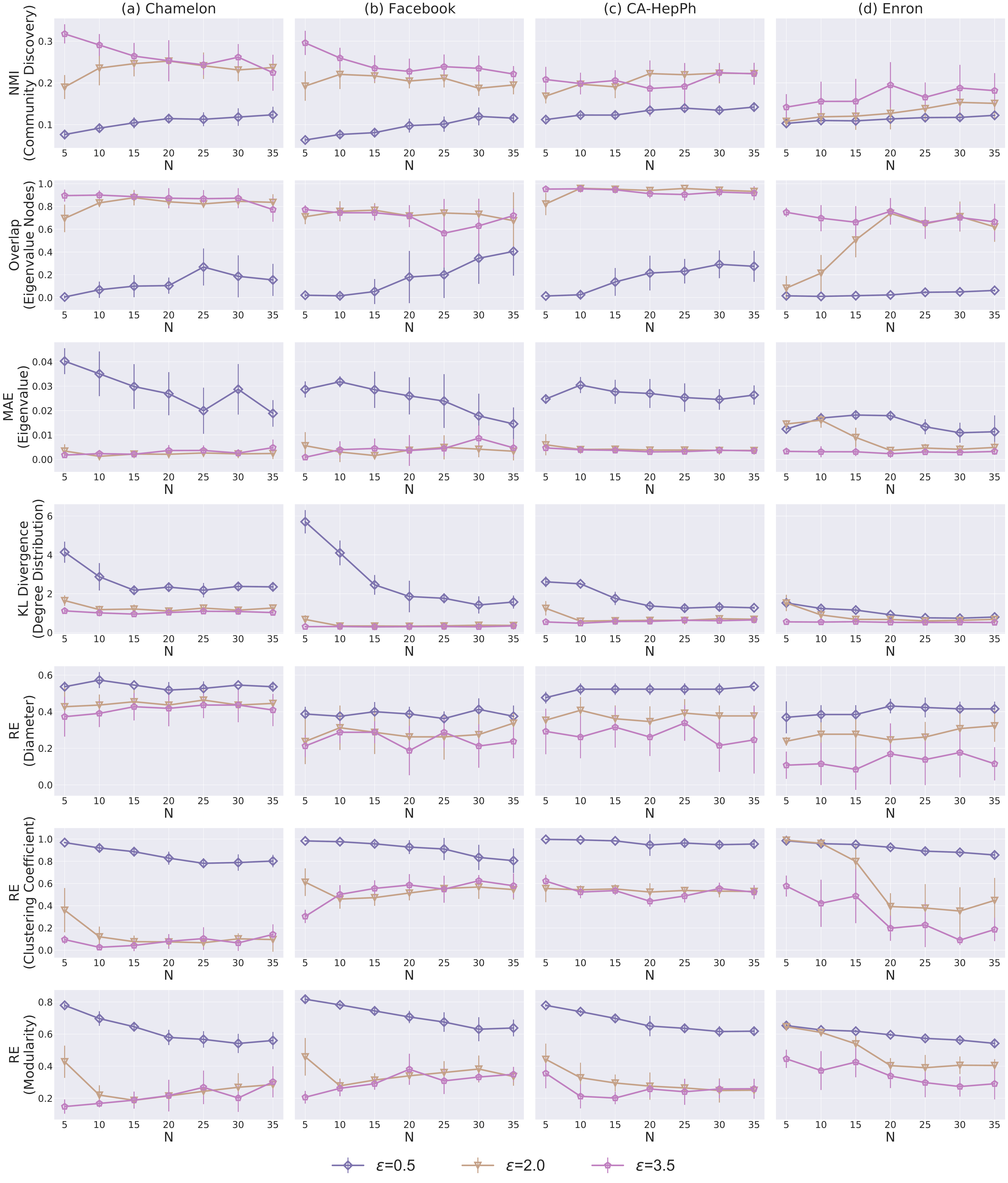}
    \vspace{-0.4cm}
    \caption{
    Impact of the number of nodes for community initialization. 
    The columns represent the used datasets, and the rows stand for different metrics. 
    In each plot, the x-axis denotes the number of nodes $N$, and the y-axis denotes the performance. 
    For the first two rows, higher is better. 
    For the last five rows, lower is better.
    }
    \vspace{-0.2cm}
    \label{fig:impact_N}
\end{figure*}

In \autoref{fig:impact_N}, 
we show the performance of \method under different combinations of $N$ and $\varepsilon$. 
When the privacy budget is small,
the noise has an significant impact on the final results,
and a larger $N$ is helpful to achieve higher accuracy because it reduces the number of initial communities.
If the privacy budget is large enough,
the injected noise has a limited effect,
and the error from random division will play a dominant role.
In this case, a smaller $N$ facilitates a better performance.

We find that there exists no fixed optimal $N$ for all privacy budgets and datasets.
For example, when the privacy budget is $0.5$,
\method usually performs better as $N$ increases.
The reason is that the privacy budget is small and the perturbation noise cannot be ignored.
When the privacy budget is up to $2$,
with the increase of $N$,
many metrics
perform better first and then worse, such as NMI, clustering coefficient, and modularity. 
This is caused by the combined effect of perturbation noise and random division error.
If the privacy budget is $3.5$,
a smaller $N$ can achieve higher results on the Chamelon and Facebook datasets because the number of communities formed is lower than the other two datasets. 
The influence of random division error is significant when the privacy budget is large.
Overall,
\method maintains competitive synthetic quality when $N=20$, thus we set $N=20$ throughout the experiments.

\begin{figure*}[!t]
    \centering
    \includegraphics[width=0.95\textwidth] {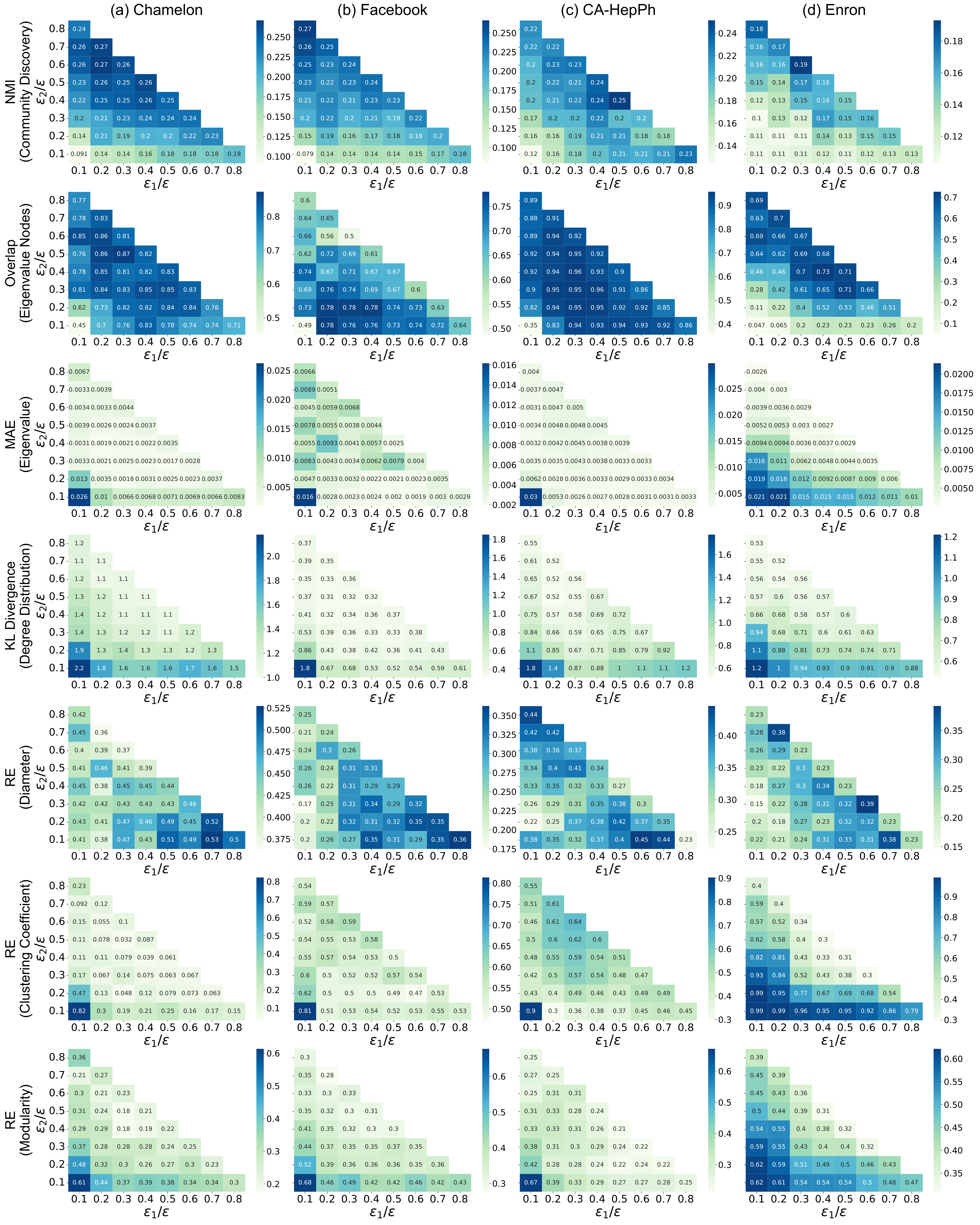}
    \vspace{-0.4cm}
    \caption{
    Impact of privacy budget allocation. 
    The columns represent the used datasets, and the rows stand for different metrics. 
    In each plot, the x-axis denotes the ratio of $\varepsilon_1$, the y-axis denotes the ratio of $\varepsilon_2$ $(\varepsilon_3=\varepsilon-\varepsilon_1-\varepsilon_2)$, and the values in the grid denote the performance.
    For the first two rows, higher is better.
    For the last five rows, lower is better.
    }
\label{fig:impact_eps}
\end{figure*}

\subsection{Impact of the Privacy Budget Allocation}
\label{subsec:privacy_budget}
Recalling \autoref{section:overview}, the entire privacy budget is divided into three parts: Community initialization ($\varepsilon_1$), community adjustment ($\varepsilon_2$) and information extraction ($\varepsilon_3$).
We vary the ratio of $\varepsilon_1$ and $\varepsilon_2$ to $\varepsilon$ from $0.1$ to $0.8$ with step size $0.1$.
Then $\varepsilon_3$ can be calculated by $\varepsilon-\varepsilon_1-\varepsilon_2$.
\autoref{fig:impact_eps} shows the impact of privacy budget allocation on four datasets and seven metrics when the privacy budget is $2$. 

We observe that the allocation ratio has an obvious influence on the experiment results.
For the overlap metric, if both $\varepsilon_1$ and $\varepsilon_2$ are small, \method cannot achieve high overlap ratios.
The reason is that the overlap ratio depends on the quality of community division.
If the communities obtained are not accurate, 
it is difficult to identify the nodes with high eigenvalues.
If the ratio of $\varepsilon_3$ is small, \method cannot obtain the best performance since the extracted information is injected lots of noise.
For the modularity metric, the behavior of \method  is poor when $\varepsilon_1$ and $\varepsilon_2$ are low, which is similar to the overlap results.
The recovery of modularity also requires accurate community partitions.
The impact of the perturbation noise from information extraction is negligible if we have proper partitions.

Based on the above observations, 
the impact of the budget allocation strategy varies on different datasets. 
Considering the universal applicability of \method, 
we set $\varepsilon_1=\varepsilon_2=\varepsilon_3=\frac{1}{3}\varepsilon$ for all experiments. 

\subsection{Impact of the Resolution}
\label{subsec:impact_resolution}

\begin{figure*}[!t]
    \centering
    \includegraphics[width=0.95\textwidth] {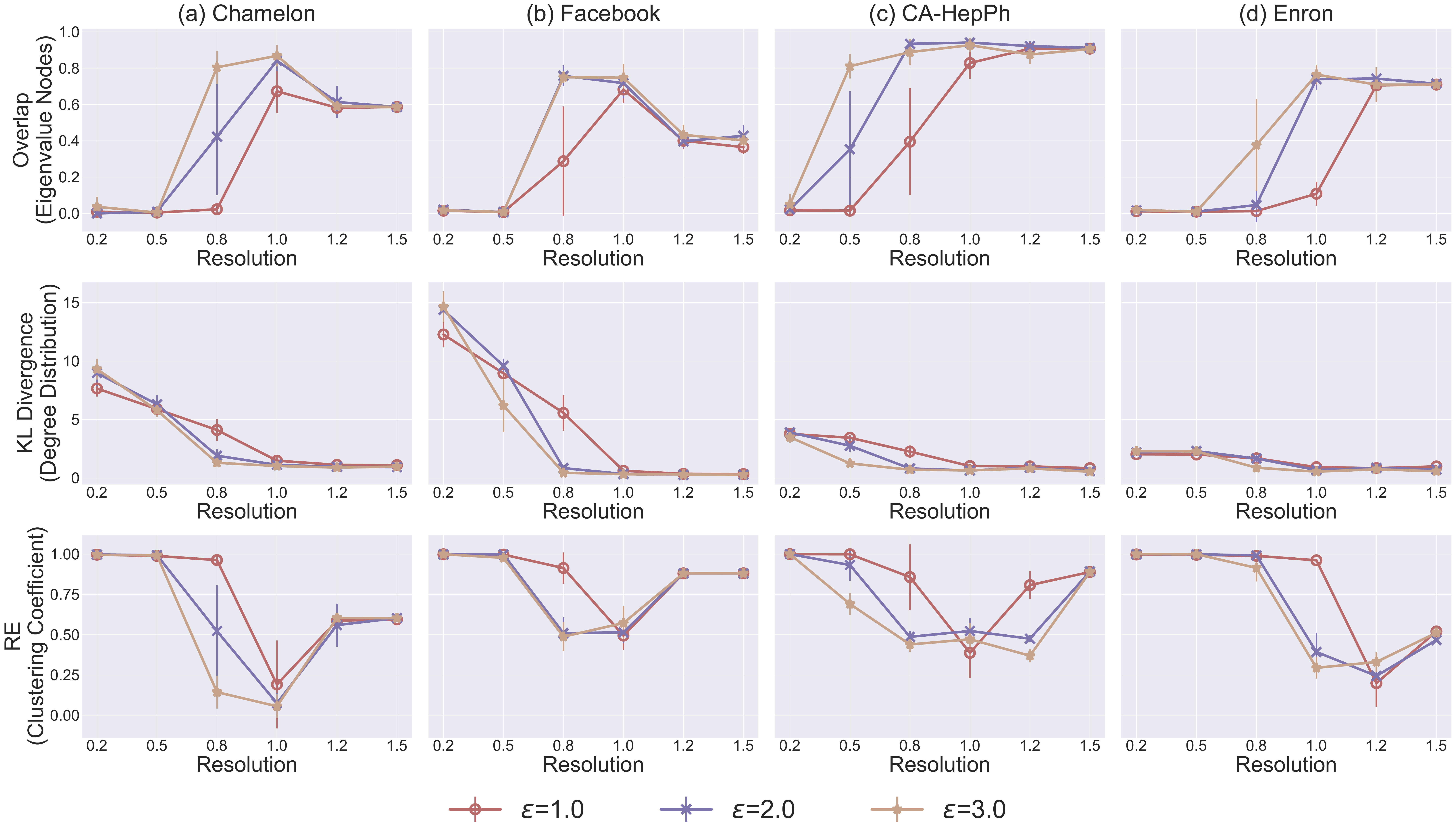}
    \vspace{-0.4cm}
    \caption{
    Impact of the resolution. 
    The columns represent the used datasets and the rows stand for different metrics. 
    In each plot, the x-axis denotes the privacy budget $\varepsilon$, and the y-axis denotes the performance. 
    For the first row, higher is better.
    For the last two rows, lower is better.
    }
    \vspace{-0.2cm}
    \label{fig:impact_resolution}
\end{figure*}

Recalling~\autoref{subsec:preliminary_community_detection}, modularity optimization may fail to detect small communities, \ie, resolution limit.
Thus, in this section, we compare the performance of \method under the resolution parameter $t$ ranging from 0.2 to 1.5. 
\autoref{fig:impact_resolution} illustrates the impact of the resolution parameter on four datasets and three metrics.

We observe that the resolution parameter $t$ has a significant impact on the final results.
For the first metric, as the resolution parameter increases, the overlap ratio shows a tendency of increasing and then decreasing.
The reason is that a too small $t$ will lead to excessive partitions, which breaks the connections between the influential nodes and other nodes, 
while the community will be insufficiently divided if $t$ is too large. 
For the degree distribution, the KL divergence shows a decreasing trend as $t$ increases since \method extracts the degree information within the community.
For the clustering coefficient, with the increase of the resolution parameter, the RE decreases and then increases.
The reason is similar to that of the overlap metric.
If the resolution parameter is too small, it will introduce a large amount of noise between the communities, and a too-large resolution parameter cannot detect the community structure accurately. 
Therefore, a suitable resolution parameter $t$ is essential to balance the effects of the above two parts.
\method shows 
consistently excellent performance when the resolution parameter $t$ is around 1.

\section{Case Study: Influence Maximization}
\label{subsec:case_study_IM}

\begin{figure*}[!t]
    \centering
    \includegraphics[width=0.95\textwidth] 
    {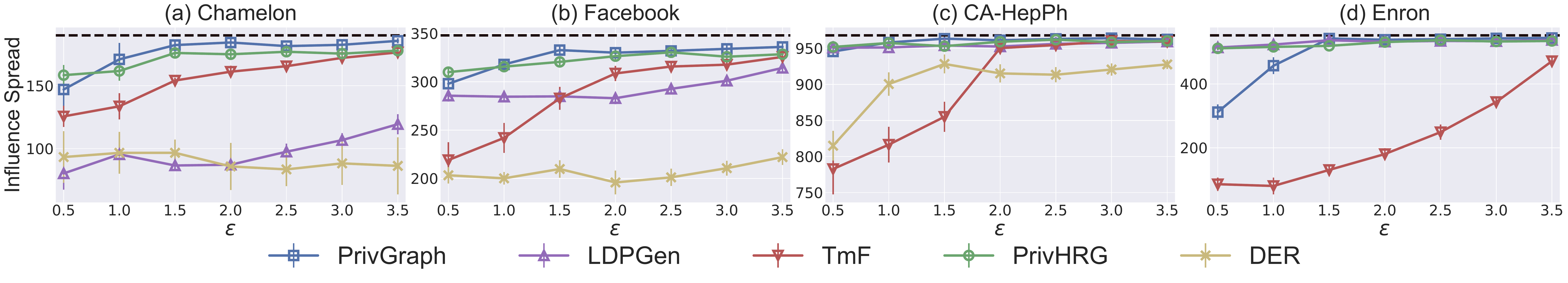}
    \vspace{-0.3cm}
    \caption{
    [Higher is better] 
    The influence spread of different methods.
    The captions represent the name of the used datasets. 
    In each plot,
    the x-axis denotes the privacy budget $\varepsilon$,
    the y-axis denotes the influence spread,
    and the black dashed line denotes the results on the original graph (without DP guarantee). 
    }
    \label{fig:Influence_Maximization}
\end{figure*}

In this section, we apply \method to the Influence Maximization (IM) problem~\cite{domingos2001mining}, which aims to locate the nodes of maximum impact from a given graph. 
A typical application of IM is viral marketing,
where a company tries to promote a new product in some targeted individuals~\cite{li2018influence}. 

\mypara{Setup}
We adopt a classical method, called Degree-Discount (DD)~\cite{chen2009efficient}, to find the top 20 of most influential nodes on the synthetic graph. 
Then, we compare the influence spreads of the selected nodes from different methods by leveraging the standard Independent Cascade (IC)~\cite{kempe2003maximizing} model with $p=0.01$ on the original graph,
where the higher influence spread values mean more accurate location of the most influential nodes. 

\mypara{Observations}
\autoref{fig:Influence_Maximization} illustrates the influence spread of the most influential nodes obtained by various methods. 
We observe that the influence spread obtained by \method is the highest in most cases. 
Recalling \autoref{section:community_division} and \autoref{section:information_extraction}, \method groups the nodes with similar connection structures into a community, and then extracts the intra-information from each community separately. 
In this way, \method recovers both the nodes' degree and graph structure, which is essential for discovering influential nodes.
When the privacy budget is limited, \eg, $\varepsilon=0.5$, the privacy budget is too small to accurately detect the communities, impacting the final influence spread. 

The performance of \gen is not stable on the four datasets.
For the last two datasets, the number of high-degree nodes is larger, which is less susceptible to noise.
The performance of \hrg is better than \tmf and \der. 
Since \hrg encodes the original graph into a tree structure and rebuilds the graph based on the connection probability between nodes, 
\hrg mitigates the perturbation impact on the original degree distributions. 
Recalling \autoref{subsec:end_to_end}, \der performs better than \hrg and \tmf on the overlap of nodes, while poorly on the influence maximization.
The reason for the inconsistency is that Degree-Discount~\cite{chen2009efficient} finds the set of seed nodes mainly from the perspective of node degree rather than eigenvalues.

\end{document}

%% file: mydefs.tex
\usepackage{epsfig,endnotes,xspace,url,diagbox}
\usepackage{amsmath,amsthm,amsfonts}

\usepackage{enumerate}
\usepackage{varioref}
\usepackage{graphicx}

\usepackage{caption}
\usepackage{subcaption}
\usepackage{float}
\usepackage{epstopdf}
\usepackage{xcolor}
\usepackage{multirow}
\usepackage{comment}
\usepackage{color}
\usepackage[utf8]{inputenc}
\usepackage{mathtools}
\usepackage{wrapfig}
\usepackage{tabularx}
\usepackage{xspace}
\usepackage[noend, ruled,linesnumbered]{algorithm2e}

\usepackage{cases}
\usepackage{stfloats}
\usepackage{fancyhdr}
\usepackage{booktabs}

\usepackage{enumitem}
\setlist[itemize]{leftmargin=*}

\hypersetup{
  colorlinks,
  linkcolor={blue!70!green},
  citecolor={green!70!blue},
  urlcolor={orange!70!red}
}

\newif\ifcomment
\commenttrue

\newcommand{\etal}{\textit{et al.}\xspace}
\newcommand{\ie}{\textit{i.e.}\xspace}
\newcommand{\eg}{\textit{e.g.}\xspace}

\newcommand{\etc}{\textit{etc.}\xspace}

\newcommand{\mypara}[1]{\noindent\textbf{#1.} \xspace}

\renewcommand{\Pr}[1]{\ensuremath{\mathsf{Pr}\left[#1\right]}\xspace}

\newtheorem{theorem}{Theorem}

\newtheorem{definition}{Definition}

\definecolor{revision}{RGB}{0,0,0}

\newcommand{\revstart}{\begin{color}{revision}}
\newcommand{\revend}{~\!\!\end{color}}

\newcommand{\method}{\ensuremath{\mathsf{PrivGraph}}\xspace}
\newcommand{\hrg}{\ensuremath{\mathsf{PrivHRG}}\xspace}
\newcommand{\der}{\ensuremath{\mathsf{DER}}\xspace}
\newcommand{\tmf}{\ensuremath{\mathsf{TmF}}\xspace}
\newcommand{\gen}{\ensuremath{\mathsf{LDPGen}}\xspace}

\usepackage{etoolbox}
\makeatletter
\patchcmd{\hyper@makecurrent}{%
    \ifx\Hy@param\Hy@chapterstring
        \let\Hy@param\Hy@chapapp
    \fi
}{%
    \iftoggle{inappendix}{%
        \@checkappendixparam{chapter}%
        \@checkappendixparam{section}%
        \@checkappendixparam{subsection}%
        \@checkappendixparam{subsubsection}%
        \@checkappendixparam{paragraph}%
        \@checkappendixparam{subparagraph}%
    }{}%
}{}{\errmessage{failed to patch}}

\newcommand*{\@checkappendixparam}[1]{%
    \def\@checkappendixparamtmp{#1}%
    \ifx\Hy@param\@checkappendixparamtmp
        \let\Hy@param\Hy@appendixstring
    \fi
}
\makeatletter

\newtoggle{inappendix}
\togglefalse{inappendix}

\apptocmd{\appendix}{\toggletrue{inappendix}}{}{\errmessage{failed to patch}}